\documentclass[sigconf]{acmart} 

\AtBeginDocument{%
  \providecommand\BibTeX{{%
    \normalfont B\kern-0.5em{\scshape i\kern-0.25em b}\kern-0.8em\TeX}}}

\copyrightyear{2022}
\acmYear{2022}
\setcopyright{acmcopyright}\acmConference[WiPSCE '22]{Proceedings of the 17th Workshop in Primary and Secondary Computing Education}{October 31-November 2, 2022}{Morschach, Switzerland}
\acmBooktitle{Proceedings of the 17th Workshop in Primary and Secondary Computing Education (WiPSCE '22), October 31-November 2, 2022, Morschach, Switzerland}
\acmPrice{15.00}
\acmDOI{10.1145/3556787.3556861}
\acmISBN{978-1-4503-9853-4/22/10}

\usepackage{xspace}
\usepackage{tabularx}
\usepackage{colortbl}
\usepackage{subfig}
\usepackage{multirow}
\usepackage{booktabs}
\usepackage{listings}
\usepackage[capitalise]{cleveref}
\usepackage{arydshln}

\newcommand\definetool[2]{\newcommand{#1}{{\textsc{#2}}\xspace}}
\definetool{\Scratch}{Scratch}
\definetool{\leila}{LeILa}
\definetool{\whisker}{Whisker}
\definetool{\litterbox}{LitterBox}
\definetool{\bastet}{Bastet}
\definetool{\scratchblocks}{scratchblocks}

\newcommand{\allfemale}{all-female~}
\newcommand{\allmale}{all-male~}
\newcommand{\mixed}{mixed~}
\newcommand{\samesex}{same-sex~}
\newcommand{\mixedsex}{mixed-sex~}

\colorlet{punct}{red!60!black}
\definecolor{background}{HTML}{EEEEEE}
\definecolor{delim}{RGB}{20,105,176}
\colorlet{numb}{magenta!60!black}

\lstdefinelanguage{json}{
    basicstyle=\normalfont\ttfamily,
    numbers=left,
    numberstyle=\scriptsize,
    stepnumber=1,
    numbersep=8pt,
    showstringspaces=false,
    breaklines=true,
    frame=lines,
    backgroundcolor=\color{background},
    literate=
     *{0}{{{\color{numb}0}}}{1}
      {1}{{{\color{numb}1}}}{1}
      {2}{{{\color{numb}2}}}{1}
      {3}{{{\color{numb}3}}}{1}
      {4}{{{\color{numb}4}}}{1}
      {5}{{{\color{numb}5}}}{1}
      {6}{{{\color{numb}6}}}{1}
      {7}{{{\color{numb}7}}}{1}
      {8}{{{\color{numb}8}}}{1}
      {9}{{{\color{numb}9}}}{1}
      {:}{{{\color{punct}{:}}}}{1}
      {,}{{{\color{punct}{,}}}}{1}
      {\{}{{{\color{delim}{\{}}}}{1}
      {\}}{{{\color{delim}{\}}}}}{1}
      {[}{{{\color{delim}{[}}}}{1}
      {]}{{{\color{delim}{]}}}}{1},
}

\usepackage{fancybox}
\newcommand{\rqsummary}[2]{
        \vspace{2mm}
        \noindent
        \fbox{%
            \parbox{.97\linewidth}{%
                    \textbf{#1 Summary.}
                #2
            }%
        }%
        \vspace{2mm}
}%

\usepackage[colorinlistoftodos,prependcaption,textsize=tiny]{todonotes}

\begin{document}

\title{Gender-dependent Contribution, Code and Creativity \\in a Virtual Programming Course}

\author{Isabella Graßl}
\email{isabella.grassl@uni-passau.de}
\affiliation{%
  \institution{University of Passau}
  \state{Passau}
  \country{Germany}
}

\author{Gordon Fraser}
\email{gordon.fraser@uni-passau.de}
\affiliation{%
  \institution{University of Passau}
  \city{Passau}
  \country{Germany}
}

 \renewcommand{\shortauthors}{Isabella Gra{\ss}l and Gordon Fraser}

\begin{abstract}	
Since computer science is still mainly male dominated, academia, industry and
education jointly seek ways to motivate and inspire girls, for example by
introducing them to programming at an early age.
The recent COVID-19 pandemic has forced many such endeavours to move to an online
setting. While the gender-dependent differences in programming courses have
been studied previously, for example revealing that girls may feel safer in
\samesex groups, much less is known about gender-specific differences in \emph{online} programming courses.
In order to investigate whether gender-specific differences can be observed in
online courses, we conducted an online introductory programming course for
\Scratch, in which we observed the gender-specific characteristics of
participants with respect to how they interact, their enjoyment, the code they
produce, and the creativity exposed by their programs.
%
%
%
Overall, we observed no significant differences between how girls participated
in \allfemale vs. \mixed groups, and girls generally engaged with the course
more actively than boys. This suggests that online courses can be a useful
means to avoid gender-dependent group dynamics.
However, when encouraging creative freedom in programming, girls and boys seem
to fall back to socially inherited stereotypical behavior also in an online
setting, influencing the choice of programming concepts applied. This may
inhibit learning and is a challenge that needs to be addressed
independently of whether courses are held online.

%
\end{abstract}



\maketitle

\section{Introduction}
\label{sec:intro}

The gender gap in computer science (CS) remains despite dedicated programs that specifically empower girls~\cite{vrieler2020}, especially in the field of software
engineering~\cite{albusays2021}. 
%
%
Programming education plays a crucial role in addressing this gap, as 
girls have different expectations, perceptions, and learning outcomes from
programming education than boys~\cite{rubio2015, vrieler2020}. This has led to
two important conclusions: First, it is important to teach programming early,
as girls' enthusiasm and self-efficacy~\cite{bandura1977} may be
better addressed at a younger age than during or after
puberty~\cite{sullivan2016}. Second, it has been suggested by many
studies~\cite{jones1995, crombie2000, margolis2002, moorman2003, lindberg2010,
zeid2011, cen2014} that \samesex courses support the engagement of
girls in CS.

In the recent past the COVID-19 pandemic has thrown a spanner in the works of many such
endeavours, forcing programming courses to move to online settings and thus
causing new challenges: Active participation is a prerequisite for good
results~\cite{gonzalez2012,zhan2015}, but it is difficult to control whether
students are actually following what they are supposed to in an online
course~\cite{hermans2017b}. Similarly, it is unclear if the virtual setting has
a negative impact on enjoyment, which is considered to be an important catalyst
for engagement~\cite{mitchell1993, teague2002, ccakir2017},
creativity~\cite{roque2016a}, or the resulting motivation~\cite{fields2014}.
Finally, the virtual setting influences how learners interact with each other
and with teachers. All of these are important factors that may influence how
girls engage with CS, thus calling for further research on gender-specific
characteristics in online programming courses.

In order to investigate possible gender-specific characteristics, we conducted
and observed an online programming course for young learners.
%
%
We designed a six-week introductory programming course in
\Scratch~\cite{resnick2009a}. In order to avoid biasing the behaviour of the
observed learners, we put a particular focus on creating only gender-neutral
tasks and exercises. We conducted this course online using a popular video
conferencing system with 33 girls and 38 boys aged 10 to 14 years, divided into
four groups: all-female, \allmale and two \mixed groups\footnote{We explicitly
do not support binary gender thinking, however, the children categorized
themselves as female and male.}. Using this setup, we investigate whether
girls' and boys' interest and interactions during the course, as well as the
code and creativity of the programs they produced as part of this course,
differ between girls and boys and between \samesex and \mixedsex groups.

Our results suggest that the virtual course setting does not reinforce
gender-dependent characteristics as a result of specific group constellations:
We observed no significant differences between how girls participated in
\allfemale vs. \mixed groups. This suggests that online courses may be a useful
means to avoid gender-dependent group dynamics. 
Comparing genders across all groups, it appears that the online setting
benefits girls: Overall, we observed that girls performed slightly better than
boys in terms of contributions, correctness of programs, creativity and
additional tasks. 
However, despite our emphasis on a gender-neutral course design, both girls and
boys resort to stereotypical topics when given creative freedom, also in an online course. This, in turn,
has an effect on which programming concepts are applied in the projects created
by boys and girls, thus potentially inhibiting further learning.
Consequently, this is a challenge that needs to be addressed independently of
whether courses are held online or in-person.

%
%


Our experiments are only the first step towards a deeper understanding of how different factors influence learners depending on gender in an online setting. In particular, further replications will be necessary in order to validate our initial findings. To support such replications and future research 
we
provide all course materials and evaluation scripts online.\footnote{\url{https://github.com/se2p/scratch-online-course}}

\section{Background and Related Work}
\label{sec:background}

\Scratch is a block-based programming language that makes basic programming concepts accessible for beginners using a simple visual format, in which they can engage creatively with programming~\cite{richard2016}. 
%
This is relevant in motivating children to learn programming, as there are gender-dependent topics that girls and boys implement differently in \Scratch projects~\cite{hubwieser2016, rubegni2020, grassl2021}. 
Similarly, the implementations and complexities of these programs have been observed to differ between genders~\cite{funke2017, grassl2021}. 

It is important to investigate these possible gender differences because the
ratio of women in CS is still comparatively low~\cite{bosu2019,albusays2021},
and a deeper understanding is a prerequisite to changing this. The possible reasons for
gender differences are manifold:
Extrinsic factors may be the persistent stereotyping of programming
as a male discipline and the few female role models~\cite{teague2002,lishinski2016}. Intrinsic reasons may be that girls often suffer from socially inherited low
self-efficacy~\cite{bandura1977}. 
Especially in the field of CS, this might lead to a tendency to seek less demanding and difficult challenges as
well as to more frustration and disinterest in the subject~\cite{betz1986,usher2008, beyer2014, redine2019,beckwith2005, fields}. 

One approach to address these problems is to divide students into \samesex groups as they may have a gender-specific positive effect
on the level of experience and positive attitude towards
computers~\cite{jones1995}. However, this is mainly observed in
first-time introductions to the subject, hence, an \allfemale group
makes sense especially in early years. This experience can be
the first major step in making girls aware of CS and their own
abilities~\cite{mitchell1993, teague2002, weese2016}. 


In the context of software engineering, grades in
\samesex classes have been observed to be slightly better for both \allfemale and \allmale groups
than in \mixed classes~\cite{zeid2011, redine2019}. Although the quality of the
 programs was similar between genders in prior studies, girls in
\allfemale classes followed the instructions more attentively, were more likely
to stick to the requirements, and submitted more programs~\cite{mcdowell2006,
zhan2015, gonzalez2012, sullivan2016}. 

%

\section{Online Course Design}
\label{sec:course}

\begin{figure}[tb]
\centering
\subfloat[\label{session1} 1: basics (story)]{\includegraphics[width=0.32\linewidth]{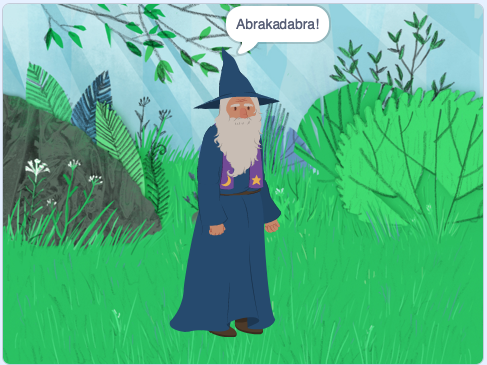}}
\hfill
\subfloat[\label{session21} 2.1: loops (anim.)]{\includegraphics[width=0.32\linewidth]{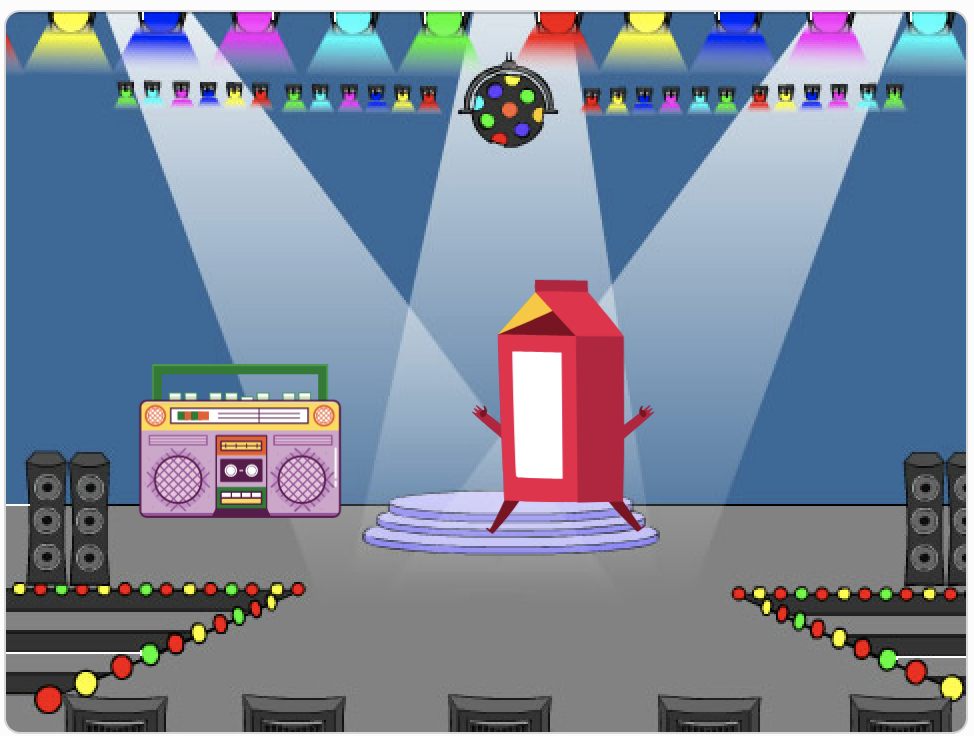}}
\hfill
\subfloat[\label{session2} 2.2: loops (story)]{\includegraphics[width=0.32\linewidth]{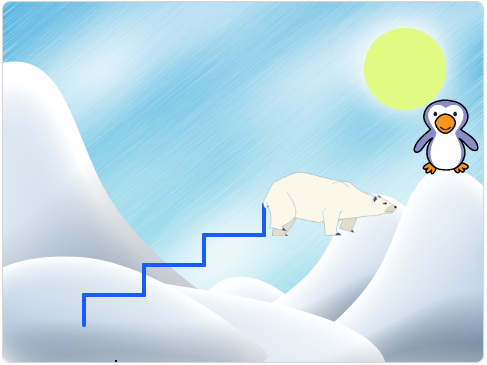}}
\hfill
\subfloat[\label{session31} 3.1: conditional statements (story)]{\includegraphics[width=0.32\linewidth]{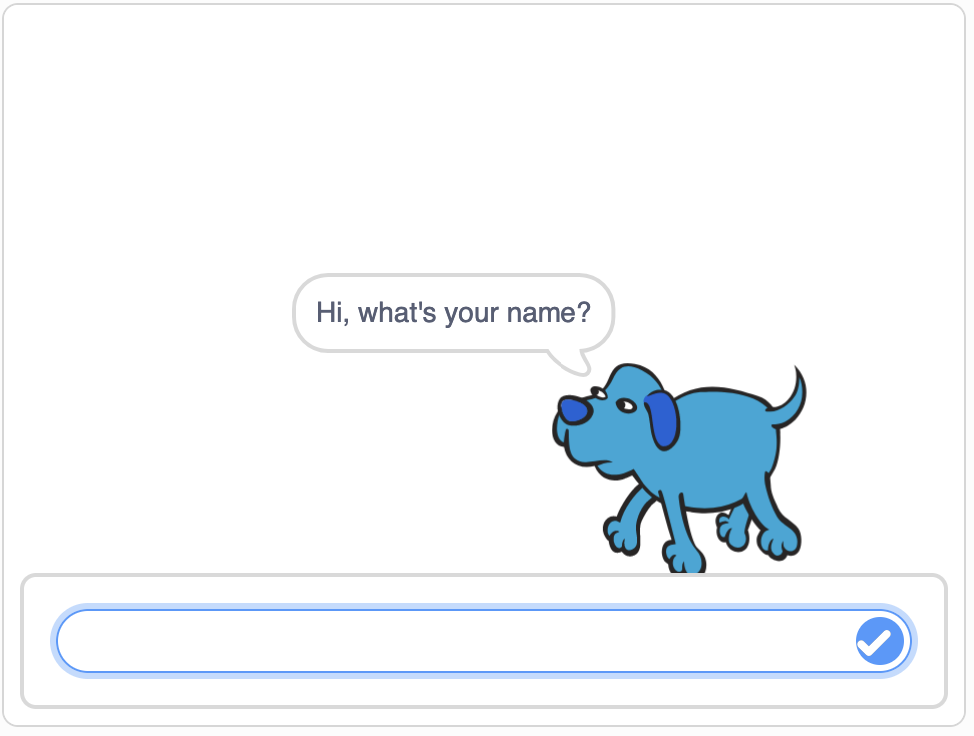}}
\hfill
\subfloat[\label{session3} 3.2: conditional statements (game)]{\includegraphics[width=0.32\linewidth]{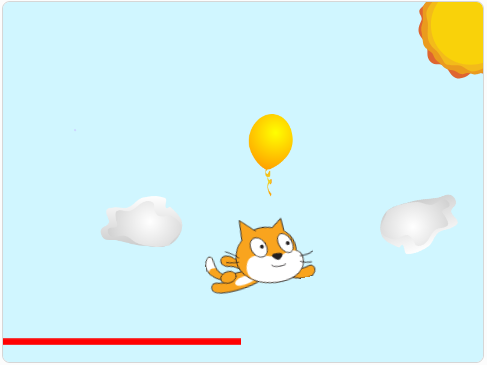}}
\hfill
\subfloat[\label{session4} 4: conditional loops (game)]{\includegraphics[width=0.32\linewidth]{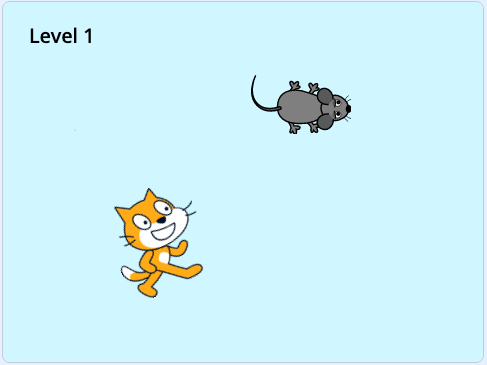}}
\hfill
\subfloat[\label{session5} 5: variables (game)]{\includegraphics[width=0.32\linewidth]{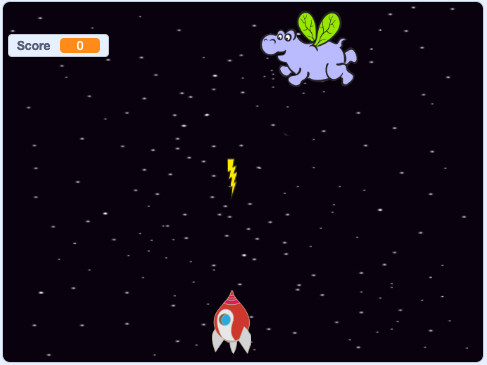}}
\hfill
\subfloat[\label{session5alt} 5: variables (game)]{\includegraphics[width=0.32\linewidth]{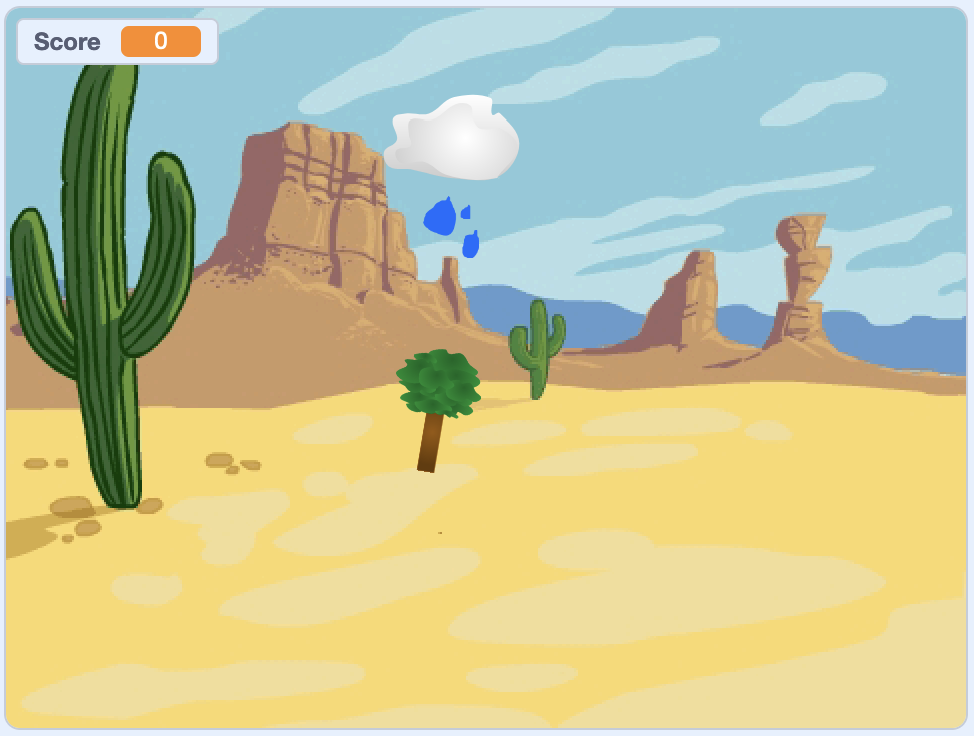}}
\hfill
\subfloat[\label{session6} 6: recap (anim./game)]{\includegraphics[width=0.32\linewidth]{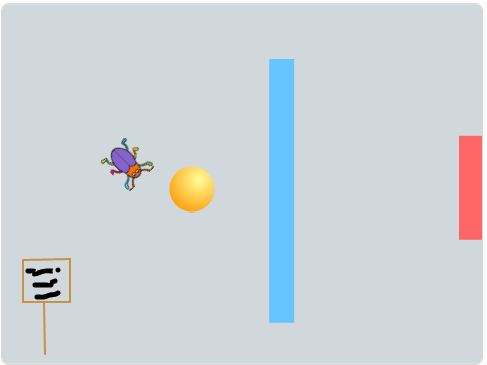}}
\caption{Design of the task for each session with the introduced programming concept and project type.}  
\label{fig:sessions}
\end{figure}

To enable our investigation, and to continue encouraging young learners to engage with programming throughout the pandemic, we created an introductory \Scratch course consisting of six one-hour lessons,
with two objectives in mind: First, the course needs to be suitable for teaching in an online, video-based scenario; second, to prevent any bias in the teaching material, all example programs and figures have to be gender-neutral as much as possible and avoid stereotypes. We provide all course material online.\footnotemark[2]

\subsection{Lesson Structure} 
\looseness=-1
In order to provide the students with a consistent learning experience, all
sessions of our course follow the same procedure: They start with a
retrospective of the previous session, which ensures that all students are up
to speed, and provides an opportunity for discussing open questions. Then, the
programming concept of the session is introduced with a sample project, which
also represents the objective of this session.
The first steps of the example program are demonstrated using the screen
sharing function of the video conferencing software, actively involving the students to work on the concrete problem solution. 
Then, the students are asked to implement the same steps by themselves.
This alternation of showing, working together and independently continues until
the programming concept of the lesson is realized, and students have co-constructed their insights.
 At the end of each session, students have the opportunity to present
their programs so that their work is valued and their self-confidence is
strengthened. 
%
After
each lesson, a PDF summary of the lesson with possible solutions is sent to the
students. 

\subsection{Online Supervision}

Each session is presented by one dedicated instructor, who is supported by at
least two additional supervisors. Students are supposed to unmute their
microphones and ask their questions directly rather than typing them, allowing other students to contribute by
answering questions. If a student has a more severe problem (e.g., technical
difficulties), the supervisors can switch to a breakout room for one-on-one supervision. 
For basic lessons, all students are kept in a single online room. For larger,
creative programming tasks after the introduction of the concept to be learned in the session (sessions 4 to 6), children are split into smaller groups with approx. 5 children and one
individual supervisor.

\subsection{Course Contents}\label{sec:contents}
The course consists of six lessons (\cref{fig:sessions}), each 
structured into sub-tasks. 
The order and
selection of programming concepts is based on the principle that sequences
represent the simplest structures, followed by loops and conditional
statements, which form the basis for conditional repetitions, which in turn
 implicitly introduces variables~\cite{grover2015,funke2017}. 
For each lesson, we provide an optional additional task as homework to further practice the concept learned.
We made sure to use
gender-neutral characters and narratives, and to find a balance between
stories and games. In particular, many animals were chosen as showcase characters because they have no gender in \Scratch and are universally popular. By doing so, we aim to avoid possible gender or cognitive
biases and allow the children the best possible level of
creativity.

\paragraph{Session 1}
The first lesson starts with questions about
programming
and the children's
experience with it as an icebreaker. 
The most basic functions of \Scratch 
are explained using a
 wizard in the forest doing magic tricks as most children like those (\cref{session1}). Based on this, the standard blocks in the \emph{looks}, \emph{sound} and \emph{motion} categories are introduced. We
also make use of the blocks of the \emph{text to speech} extension, which allows a more unique media experience~\cite{fields2014}. Only sequences are used which are perceived as being easy for beginners~\cite{grover2015}. The transformation of the wizard into a rabbit is intended as additional task.

\paragraph{Session 2}
The programming concept \textit{loops} (\texttt{forever},
\texttt{repeat}-\texttt{times}) is introduced using a `dancing' milk carton, where the
dancing is simulated by switching costumes, while standing on a stage with
music from a music box (\cref{session21}). This example clearly demonstrates the use for loops as
it quickly leads to a feeling of success, and additionally the choice of music
provides a degree of creative freedom. The milk sprite is one of the few characters in the standard catalogue that make the appearance of dancing when changing costumes, but does not have a gender and does not fall into stereotypes. 
In a second task (\cref{session2}), the loop concept is practiced with a polar
bear drawing a number of steps up to a penguin and saying \textit{hi}.
In an additonal task, the polar bear should do a `flip' after each step, using a
nested loop inside the outer loop.

\paragraph{Session 3} 
\textit{Conditional statements} (\texttt{if}, \texttt{if}-\texttt{else}) are
introduced with a dog that asks the students what their name
is (\cref{session31}). The students are immediately involved by choosing their own figure to ask for their names, thus
establishing a lifeworld connection~\cite{fields2014}. As a second task,
a simple game in which a flying cat is controlled
using the cursor keys, and has to fly to the sun without touching any clouds,
is created. As an extra task, this program can be
extended with multiple clouds, balloons that increase the speed, or a
background with a winning-message.

\paragraph{Session 4} (\cref{session4})
In the fourth lesson, conditional repetition (\texttt{repeat}-\texttt{until}) is introduced because it
presupposes the basic concepts of loops and conditional statements. 
We designed a game in which the cat has to catch the
mouse. The cat is controlled by the cursor keys and moves continuously until it
catches (i.e., touches) the mouse, and then says ``gotcha''. As an additional
task, different levels can be included using backdrop changes, for example, a
chick asks the cat if it can turn it into a butterfly. Implicitly, variables
are already present in the conditional repetition, which the next lesson
explicitly addresses with active settings and changes~\cite{grover2015}.

\paragraph{Session 5} 
For the introduction of variables, we created two alternative versions of the
same game for which students can vote which version they want to see. The game
either consists of flying hippos that can be zapped with lightning
(\cref{session5}), or trees that can be watered with raindrops
(\cref{session5alt}). In both cases, variables are used to track the player
score. Both games have the same code, but we
explicitly wanted to show a positive event handling with the enlivening of a
tree and one event with the destruction of something, but in a child-friendly
way. 
As an additional
task, the game can be extended with different levels, or by asking the name of
the player and storing it in a variable before starting the game.

\paragraph{Session 6} (\cref{session6})
The final lesson starts with a recap of all concepts learned using a 
program containing all the figures from the previous sessions. The task for
students is then to apply all these concepts in their own final program, in
which they create different backdrops representing rooms or levels (variables)
into which their chosen character is moved through doors
controlled by the cursor keys. As additional task, they should creatively expand their
world in some way, for example, adding signs as hints or figures 
that they should avoid. The demonstration project consists of a ball
moving through colored doors on a gray background, dodging a beetle and
receiving hints from signs. This choice aims to avoid that the children are
influenced by the example project.

\section{Method}
\label{sec:method}

In this paper we aim to empirically answer the following research questions in the context of an online programming course:
\begin{itemize}
\item[\textbf{RQ1:}] \textit{Do girls and boys show different levels of contribution?} 
\item[\textbf{RQ2:}] \textit{Do girls and boys show different levels of interest?}
\item[\textbf{RQ3:}] \textit{Do programs of girls and boys show differences in code?} 
\item[\textbf{RQ4:}] \textit{Do programs of girls and boys show differences in creativity?}
\end{itemize}

\subsection{Data Collection}

We conducted the synchronous online course in April/May 2021 over six
weeks, targeting children aged 10 to 14 without prior programming
experience. An open invitation to the course was posted on the
university and department websites, a local newspaper, and it was sent
to all schools in the area by mail. We explicitly encouraged girls to get
involved with programming in the
invitation by asking them to simply give programming a try. 
The course was conducted online using the video conferencing software
Zoom\footnote{https://zoom.us}. Although each session was scheduled for 60 minutes, this time
was usually exceeded 
due to ongoing questions. We used Zoom's survey feature to collect
survey data at the end of each session.
%
%
Researchers filled different roles within the sessions to ensure a clear course
flow and data logging: an instructor explained the tasks and led through the
course, an observer logged all interactions in a protocol, and at least two
supervisors were available for questions, of which they also kept a log. The
leadership roles were gender-balanced so that three sessions were led by a
woman and three sessions were led by a man. We used the \Scratch classroom feature and created a classroom and user accounts for participants, such that we had access to the projects created by the participants during and outside the lessons.

\subsection{Dataset}
A total of 88 students enrolled in the course, of which 8 did not show up and
80 participated. Out of these 80, we isolated 9 students who revealed prior
experience up front into a separate ``advanced'' group, which is not included
in the analysis. We consider in the analysis 71 German students
(33 female, 38 male) randomly distributed by their sex in four groups, aged 10 to 14 years (age/sex: A=all-female: 10.68/16f, B=all-male: 11.35/18m, C=mixed: 11.00/10f/8m, D=mixed: 11.73/8f/11m).
%
The majority of students attended high school (71,62 \%), although students from
junior high (14,86 \%) and elementary schools participated (13,51 \%). All sessions were consistently very well attended by all
students~(\cref{tab:studentSession}). 
Besides the protocols written by the researchers, the dataset consists of
demographic and sex information for the participants (elicited as part of the
course registration), survey responses collected at the end of each session via
Zoom, and the participants' \Scratch projects. We accessed and downloaded all
projects created by the participant users in the \Scratch classroom six weeks
after the end of the course.

\begin{table}[tb]
\centering
\small
\caption{Attendance of the students per session and in total.}
\label{tab:studentSession}
\begin{tabular}{llrrrrrrr} 
\toprule
 Group  & Sex &  \multicolumn{6}{c}{Session}   & Total  \\
 	  & & 1  & 2  & 3  & 4  & 5  & 6  & 	  \\
 \midrule
A (f)  &f& 13 & 14 & 15 & 15 & 14 & 12 & 16     \\ \hdashline
B (m)   &m& 16 & 17 & 16 & 17 & 15 & 15 & 18     \\ \hdashline
\multirow{2}{*}{C (f/m)} & f & 7 & 9 & 9 & 8 & 8 & 7 & \multirow{2}{*}{18} \\
 & m & 8 & 7 & 7 & 8 & 7 & 7 &  \\ \hdashline
\multirow{2}{*}{D (f/m)}& f& 7  & 8  & 8  & 7  & 8  & 6  & \multirow{2}{*}{19}     \\
 & m & 10 & 10 & 9  & 8  & 9  & 11 &     \\
\bottomrule
\end{tabular}
\end{table}

\subsection{Data Analysis}


\paragraph{\textbf{RQ1: Contribution}}
In order to determine the contributions of individual participants throughout the sessions, we noted all interactions in written protocols for the main Zoom session as well as the breakout groups. 
Each
observation was assigned an ID, the student anonymized, and sex marked. A
coding scheme was designed and each observation was labeled using these
different categories and subcategories. Participation focused specifically on
the categories of question and answer and whether they were \Scratch-related,
organizational, technical, or private. In addition, reactions and contributions
that are considered smalltalk were categorized. To answer RQ1, we compare the
questions and answers provided dependent on gender and group constellation.

\paragraph{\textbf{RQ2: Interest}}
In order to determine the subjective perceptions of the students, we conducted a
short survey at the end of each session, and evaluate the responses for RQ2.
Each survey included three 3-point Likert scale questions about whether (1)~the
lesson was fun and comprehensible (2)~how difficult the lesson was, and
(3)~whether the students engaged with \Scratch at home. The results of the
surveys were aggregated, anonymized, and gender-tagged.

\paragraph{\textbf{RQ3: Code}}
We evaluate the code of the students' programs using
\litterbox~\cite{fraser2021litterbox} to extract statistics about size, complexity, and other code attributes.
In addition, we manually verified for all sessions and participants whether the
concepts learned were implemented correctly and whether the additional tasks were
completed. We reviewed to what extent the students published their
projects, and which projects they created at all. We compare these
statistics in terms of gender and group context, and measure statistical
differences using a Wilcoxon Rank Sum test with $\alpha=0.05$.

\paragraph{\textbf{RQ4: Creativity}}
We evaluate the content of the students' programs implemented in the
course. We define creativity by manually checking whether any creative
deviations exist from the sample projects, such as custom sprites and
backgrounds, or additional figures and colors.
Each program was reviewed by two independent researchers, resulting in an inter-rater agreement of $K$ =
0.92. We compare the number of creative deviations and qualitatively study the topics chosen.

\subsection{Threats to Validity}

\looseness=-1
Due to the virtual setting, 
 we cannot always ensure that children worked alone.
Parents consented to use the data for research purposes and no image or audio files of the children were recorded.
 The observations
in the course are subjective in nature, but these come from three independent
supervisors per session. The evaluation of the protocols and projects
can lead to misinterpretations, so these were also evaluated by several
independent researchers. 
The design as well as the
realization of the introductory course were
evaluated by several independent didactic experts.
Students were all from Germany with no further data available on social
background. 
While the number of
participants was large relative to the desired target group size for our
course, more datapoints will be needed for generalizable results. In particular, the number of supervisors as well as observers limits the number of courses that can be reasonably conducted. Although we use the common value of $\alpha=0.05$ for statistical tests, observations are of interest independently of this arbitrary boundary, so we provide the raw $p$-values to allow the readers to better interpret the data. We therefore
provide all material and hope that this enables future replications of our online study.



\section{Results}

\begin{table}[t]
\centering
\caption{Comparison of metrics between \samesex and \mixed groups by their statistical significance ($a = 0.05$).}
\label{tab:correlations}
\begin{tabular}{llr@{\hspace{0.9em}}r@{\hspace{0.9em}}r@{\hspace{0.9em}}r@{\hspace{0.9em}}r}
\toprule
RQ & Metric & A-B & A-C(f) & A-D(f) & B-C(m) & B-D(m) \\
 \midrule
RQ1& contribution & .06 & .43 & .68 & .31 & .09 \\
RQ2& fun & .44  & .38  & .44 & .27 &  .41 \\
& difficulty & .24 &  .49 & .40 &  .29 & .74 \\
& engagement & .08  & .49 & .18  & .46 & .19  \\
RQ3& correct tasks & .10 & .94 & .01 & .25 & .46 \\
& add. task & .15 & .43 & .68 & .43 & .84 \\
RQ4& add. sprite & .06 & .49 & .99 & .64 & .07 \\
\bottomrule
\end{tabular}
\end{table}

Across all research questions, we observed almost no significant
differences between the individual \samesex and \mixed groups, i.e.,
comparing data between girls in the \samesex and \mixed groups, and
comparing data between boys in the \samesex and \mixed groups. The
$p$-values are summarized in \cref{tab:correlations}: There is only a
single significant difference, comparing the number of correct tasks
(RQ3) between girls in the \samesex and \mixed group D. We conjecture
that this $p$-value is influenced by the relatively small number of
girls in group D, who happened to provide a lower number of
submissions in the last two tasks. Thus, overall we conclude that
same-sex groups seem to provide no benefits in our online setting.

\rqsummary{RQ1--4 Same-sex vs. Mixed Groups}{We observed almost no significant
differences between how girls participate and perform in \samesex and \mixed
groups in an online setting, and there are also no significant differences how
boys participate in \samesex and \mixed groups.}

In the following we will therefore only discuss the overall differences between
girls and boys across all groups, and omit further comparisons of groups.

\subsection{RQ1: Contribution}
\label{sec:rq1}

\begin{table}[tb]
\centering
\caption{Categories of the observations for each group.}
\label{tab:protocolCategory}
\begin{tabular}{llrrrr}

\toprule
Category & Subcategory &  A (f)  &   B (m)  &   C (f/m)  &  D (f/m)   \\
\midrule
answer & correct &  51 &  56 &   64 &   47  \\
         & incorrect &   5 &   7 &   18 &    4  \\
         & suggestion &   4 &   1 &    1 &    6 \\
chitchat & general &   2 &   9 &   10 &    0 \\
         & task-related &   8 &   8 &   11 &   25 \\
question & creativity &   9 &   9 &    6 &   13 \\
         & organizational &   6 &   7 &    7 &    7 \\
         & scratch general &  16 &  31 &   13 &   10 \\
         & task-related &  83 &  87 &  109 &  108 \\
         & technical &   4 &   3 &   15 &    6  \\
reaction & status &  14 &   3 &   10 &    9\\
         & showing project &  15 &  17 &   14 &   18  \\         
         & other &   7 &   4 &    7 &    5  \\
$\sum$ &  &  230 &   242 &   285 &    258 \\
\bottomrule

\end{tabular}
\end{table}



\begin{figure}[tb]
\centering
\includegraphics[width=\columnwidth]{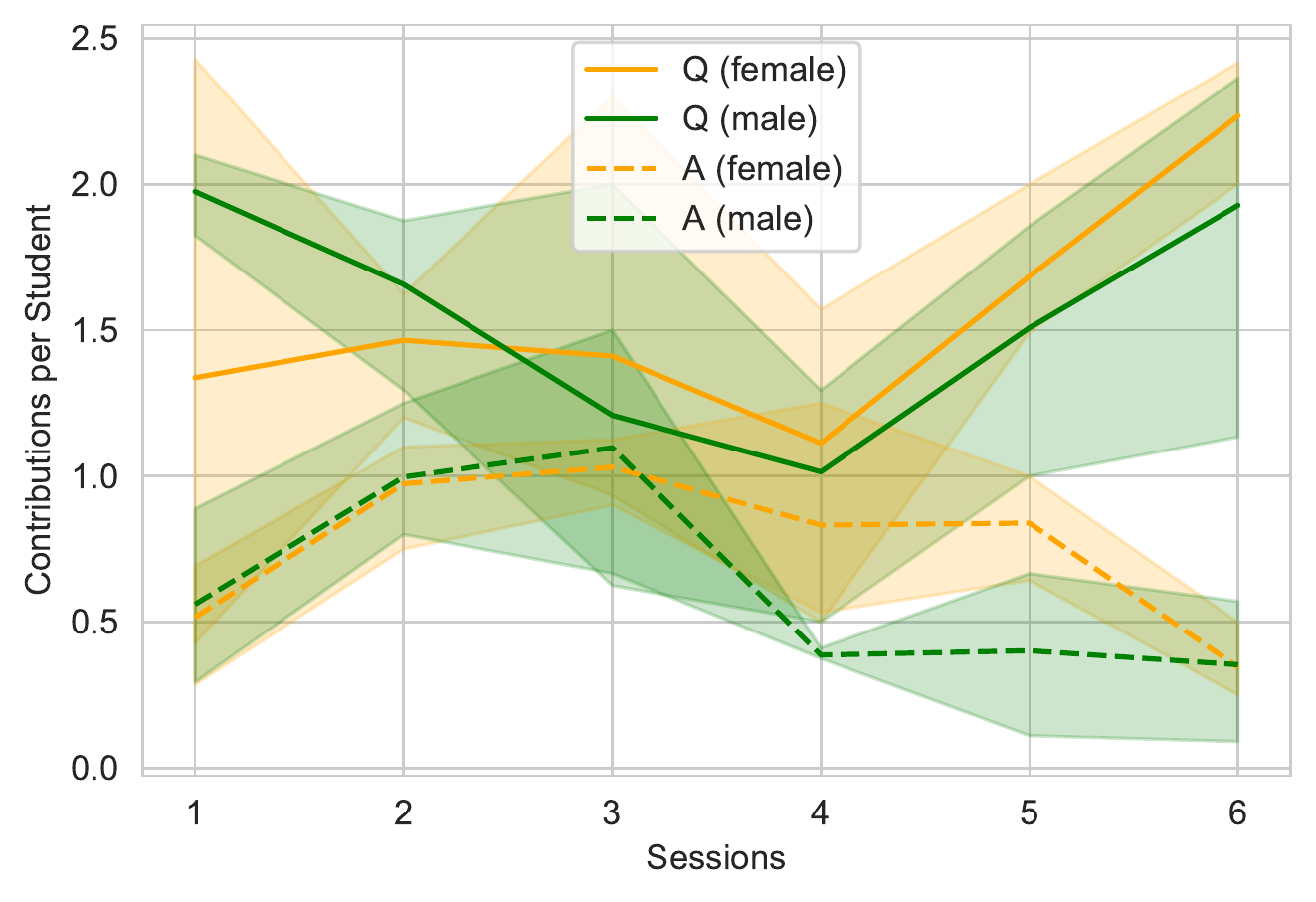}
\caption{Questions and answers of students.}
\label{fig:qa}
\end{figure}

\begin{figure*}[tb!]
\centering
\subfloat[\label{fun}Fun.]{\includegraphics[width=0.32\linewidth]{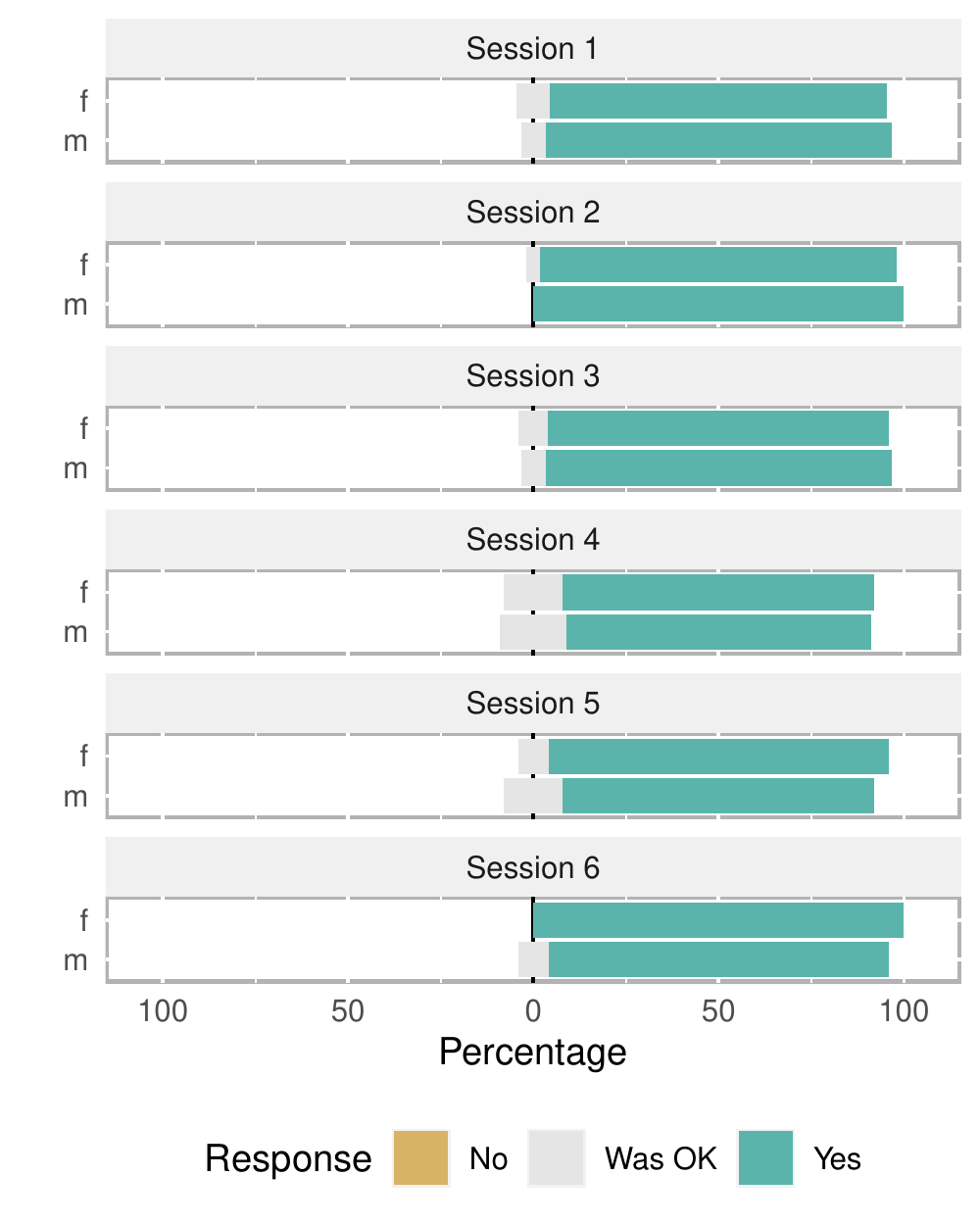}}
\hfill
\subfloat[\label{difficulty}Difficulty.]{\includegraphics[width=0.32\linewidth]{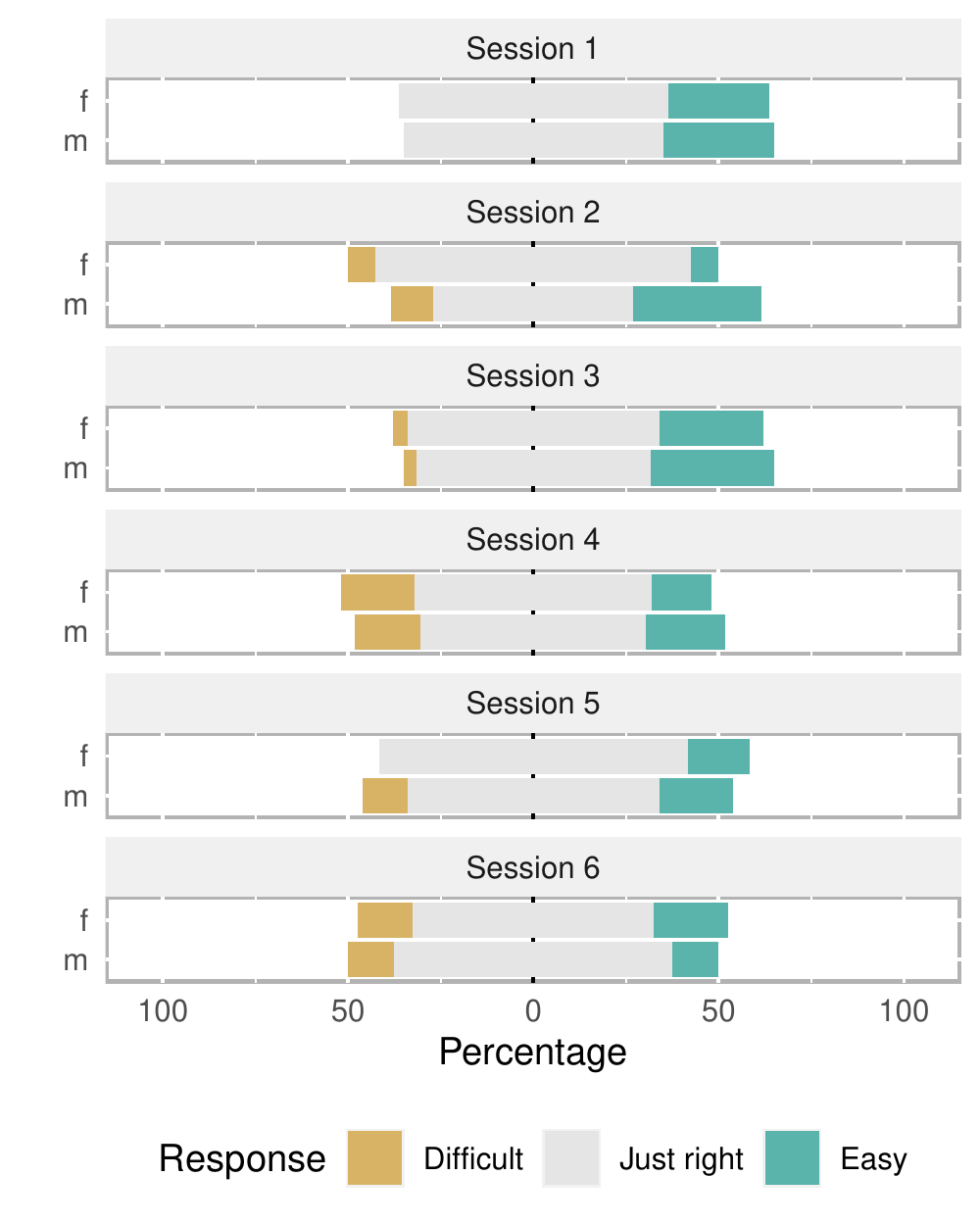}}
\hfill
\subfloat[\label{engagement}Engagement outside class.]{\includegraphics[width=0.32\linewidth]{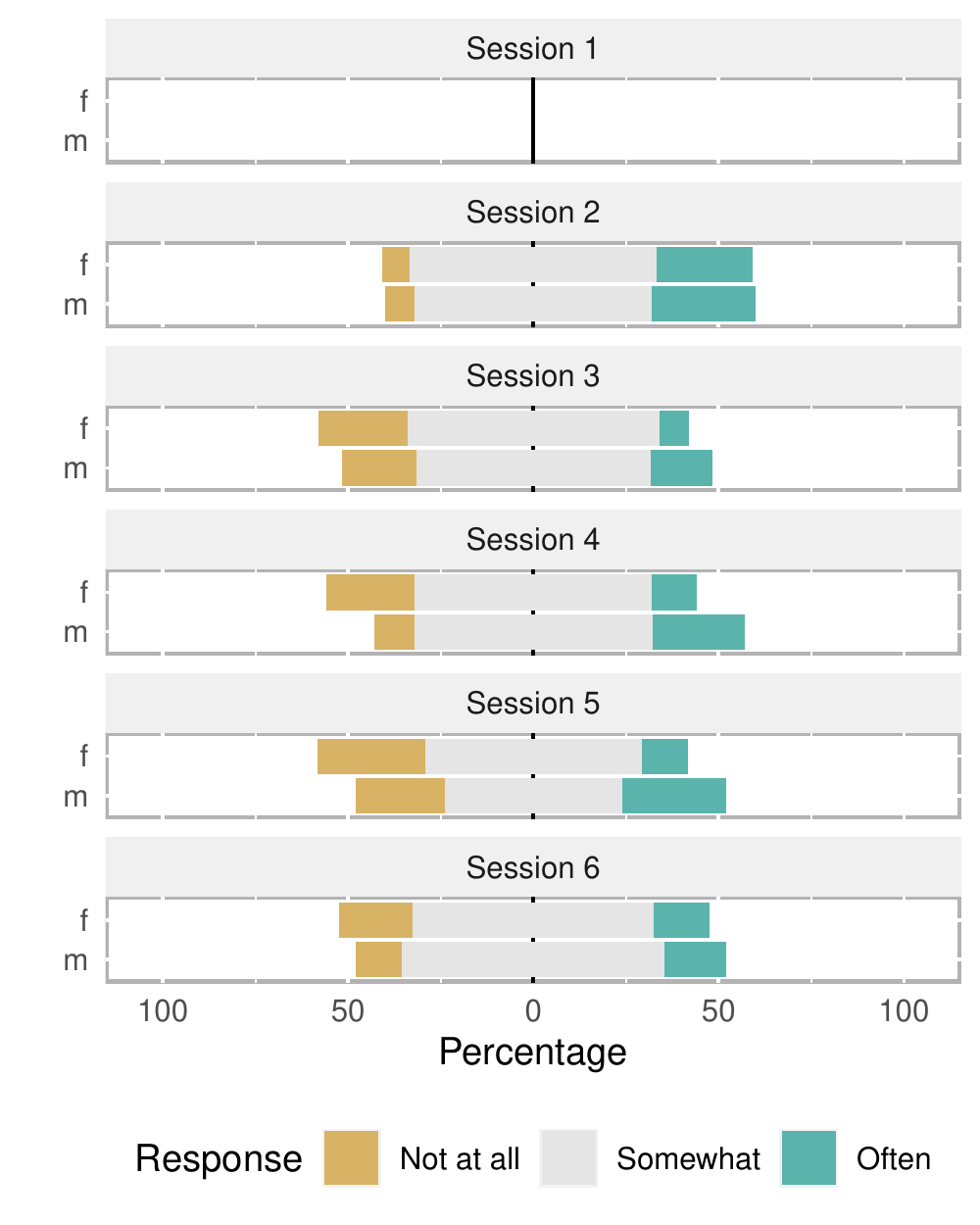}}
\caption{\label{fig:survey}Distribution of fun, difficulty, and engagement.}
\label{fig:interest}
\end{figure*}

Overall, almost across all groups and the
whole course, over 50 \% of all students actively contributed---with the
highest active participation of all groups in the fourth session with over 80
\%. For a virtual setting in particular, this is significant as it shows that
students were involved~\cite{gonzalez2012, hermans2017b}.
\Cref{tab:protocolCategory} breaks down the observations into the categories and subcategories for each group. Overall, the two \mixed groups (C, D) have the most active participation, while the two \samesex groups (A, B) have comparable values in the frequency of participation.

\Cref{fig:qa} illustrates the contributions classified in questions (solid
lines) and answers (dashed lines) in total, summed up and divided by the
participants per session. The plot further shows the 95 \% confidence
intervals; intuitively, if the confidence intervals overlap, there is no
statistically significant difference. The total number of questions and answers over all sessions is comparable between female and male students (f: 407, m: 435). In terms of a virtual setting, the number of contributions on average per session regarding questions and answers is striking (f: 67.83, m: 72.50). 

%
%
%

According to \Cref{fig:qa}, more questions were asked than answers
given regardless of gender, which is due to the nature of an
introductory course. For both genders, the highest number of questions
as well as the highest ratio of questions to answers can be seen in
the first and last sessions: In the first session mostly
organizational matters and technical problems were discussed and there
was not much room for students providing their own answers. In the
last session, no new programming concepts were introduced, so there
was intuitively less need for answers, but because students were
encouraged to integrate their own ideas many questions were raised.
The ratio of questions and answers is most balanced in the second and
third sessions. This may be due to the course design, as in these two
sessions an introductory task was implemented before moving on to the
main task, where it consequently may have been perceived as easier for
the students to participate.
In the fourth session participants of both genders showed the least
interactions, which may be related to the complexity of the
programming concepts introduced in this session (conditional loops).
%

Comparing the behavior of girls and boys, especially
in the first session, boys asked about one third more questions than girls~(\cref{fig:qa}).
However, girls participated more often than boys from the third session onwards
with questions as well as answers, which may be due to their initial lower
self-efficacy or self-confidence~\cite{lishinski2016,weese2016}. Increased
confidence after positive experiences apparently caused them to participate
more actively overall~\cite{beckwith2005,ccakir2017}. In addition, the high
response rate in sessions two and three may suggest that girls mainly respond
to quick positive feedback~\cite{vrieler2020} that they received through the
introductory tasks in these sessions. Notably, girls provided significantly more answers in the fourth session, which introduced more advanced programming concepts~(\cref{fig:qa}).

\rqsummary{RQ1}{All students contributed very actively. Girls asked fewer questions in the first sessions, but contributed considerably more overall than boys after the third session.}

The virtual setting has the potential to avoid gender-dependent group dynamics. Quick positive feedback in tasks and first positive experiences seems also relevant for a higher contribution in an online course, especially for girls. 

\subsection{RQ2: Interest}
\label{sec:rq2}

%
\Cref{fig:survey} summarizes the results of the surveys conducted at the end of each sessions.
Both genders overwhelmingly enjoyed the course (f: 92~\%, m:~91~\%;~\cref{fun}).
The majority of students also felt the difficulty of the sessions
(\cref{difficulty}) was appropriate (f: 73~\%, m: 65~\%), with boys finding the
sessions slightly easier (f: 19~\%, m: 26~\%). Overall, this shows that the
course was appealingly designed and realized.
%
%
The fourth session stands out slightly with somewhat lower reported fun and
higher reported difficulty. This may be related both to the greater
independence of the increasingly free tasks and to the flattening learning
curve, so that the advanced programming concepts are more challenging to
acquire. 
%
%

According to their self-reported engagement (\cref{engagement}), the majority of participants of both genders engaged \textit{somewhat} with \Scratch individually after the sessions (f: 64~\%, m: 62~\%). 
The claimed engagement following the second session stands out with both genders engaging \textit{often} with \Scratch almost similarly, which may be due to their initial motivation and curiosity. From sessions 3 onwards, boys claimed to engage outside the classroom more than girls. This engagement, however, is not reflected in completion rate of the additional task of session two, which will be discussed as part of RQ3 (cf. \cref{fig:addtaskcomparison} in \cref{sec:rq3engagement}). 
%



\rqsummary{RQ2}{Both genders enjoyed the course and engaged equally. The majority found the difficulty to be adequate.}

Since the virtual setting did not seem to seem to have any negative effects on the students' interest of either gender, it is suitable for further courses. The small differences observed suggest that teachers should maybe try to strengthen the girls' self-concept due to the difference between it and their actually performed tasks.

\subsection{RQ3: Code}
\label{sec:rq3}

In order to answer RQ3, we consider the code of the \Scratch projects created by the learners.


\begin{figure}[tb!]
\centering
\includegraphics[width=\columnwidth]{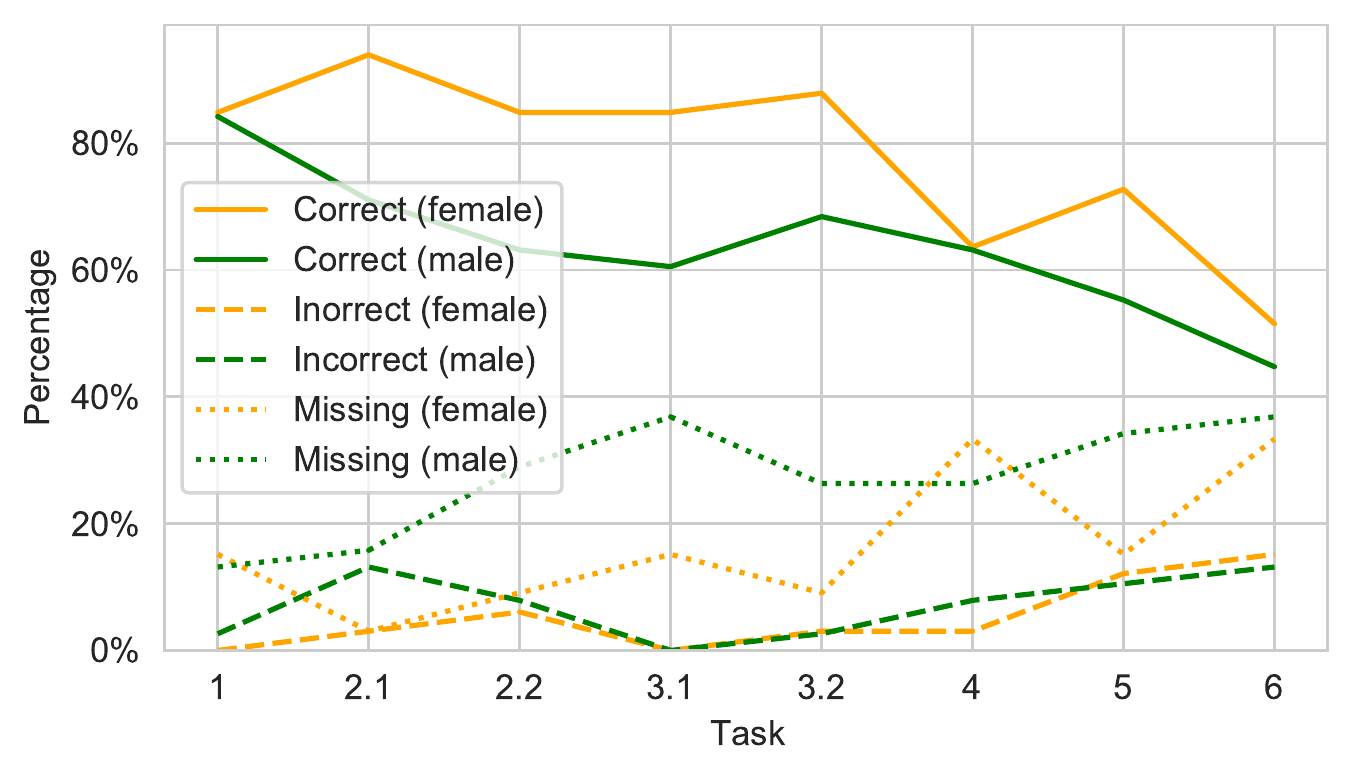}
\caption{Distribution of implementing the tasks correctly.}
\label{fig:taskcorrectsex}
\end{figure}

\subsubsection{Correctness} 


\Cref{fig:taskcorrectsex} summarizes to what degree the individual tasks were correctly solved by the participants throughout the course, separated by gender.
Girls provided correct implementations of the
tasks considerably more often than the boys over the entire course ($p =
0.18$), with particularly pronounced differences in the second and third
session.
Both genders also show a similar pattern of incorrect implementations, but the percentage is higher for boys ($p = 0.16$).
Boys' projects are also completely missing more often than girls' projects
($p = 0.12$), in total almost twice as many are missing from boys (83) than from girls (44). 
This generally suggests that girls appear to follow the tasks more attentively,
are less distracted and more into the subject because their projects are less
absent, which is in line with the findings of recent studies~\cite{mcbroom2020}.


\begin{figure}[tb]
\centering
\includegraphics[width=\columnwidth]{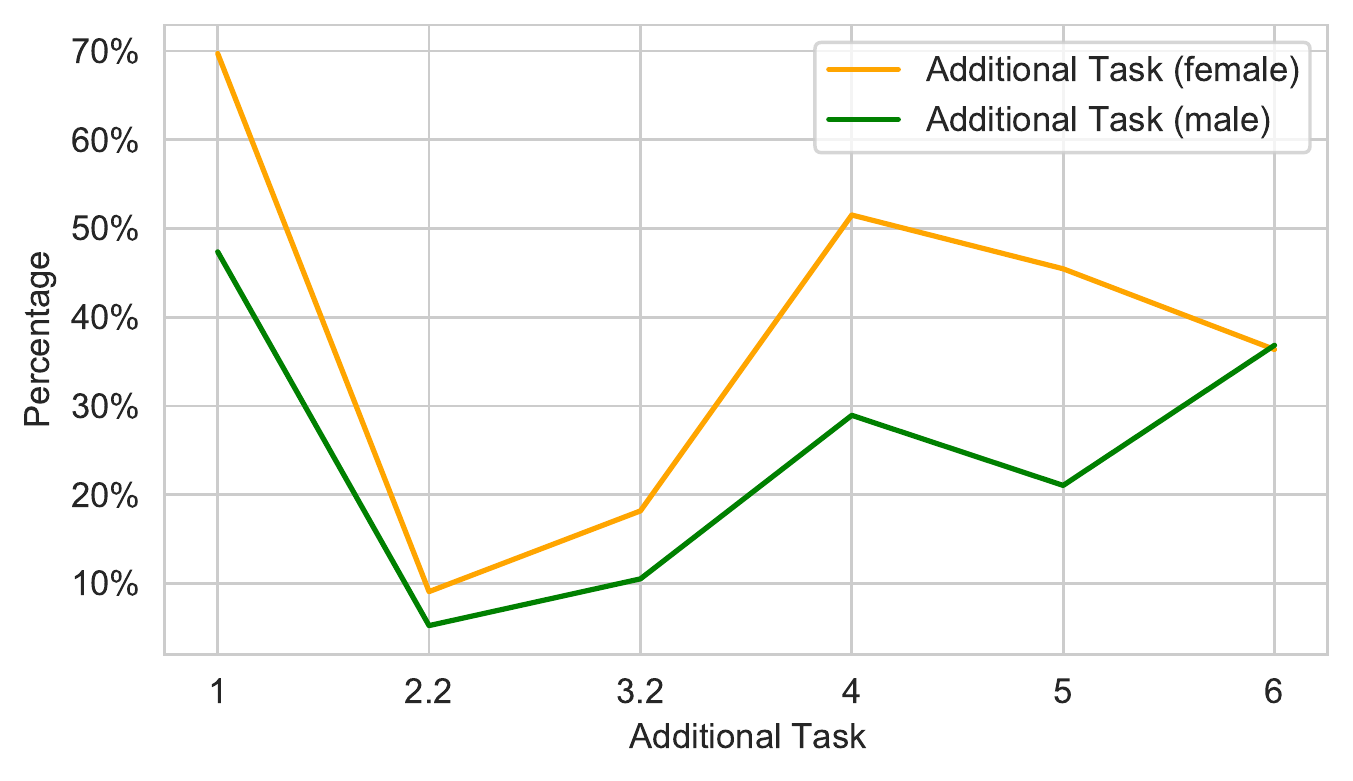}
\caption{Distribution of implementing additional tasks.}
\label{fig:addtaskcomparison}
\end{figure}

\begin{figure}[tb]
	\centering
	\includegraphics[width=\columnwidth]{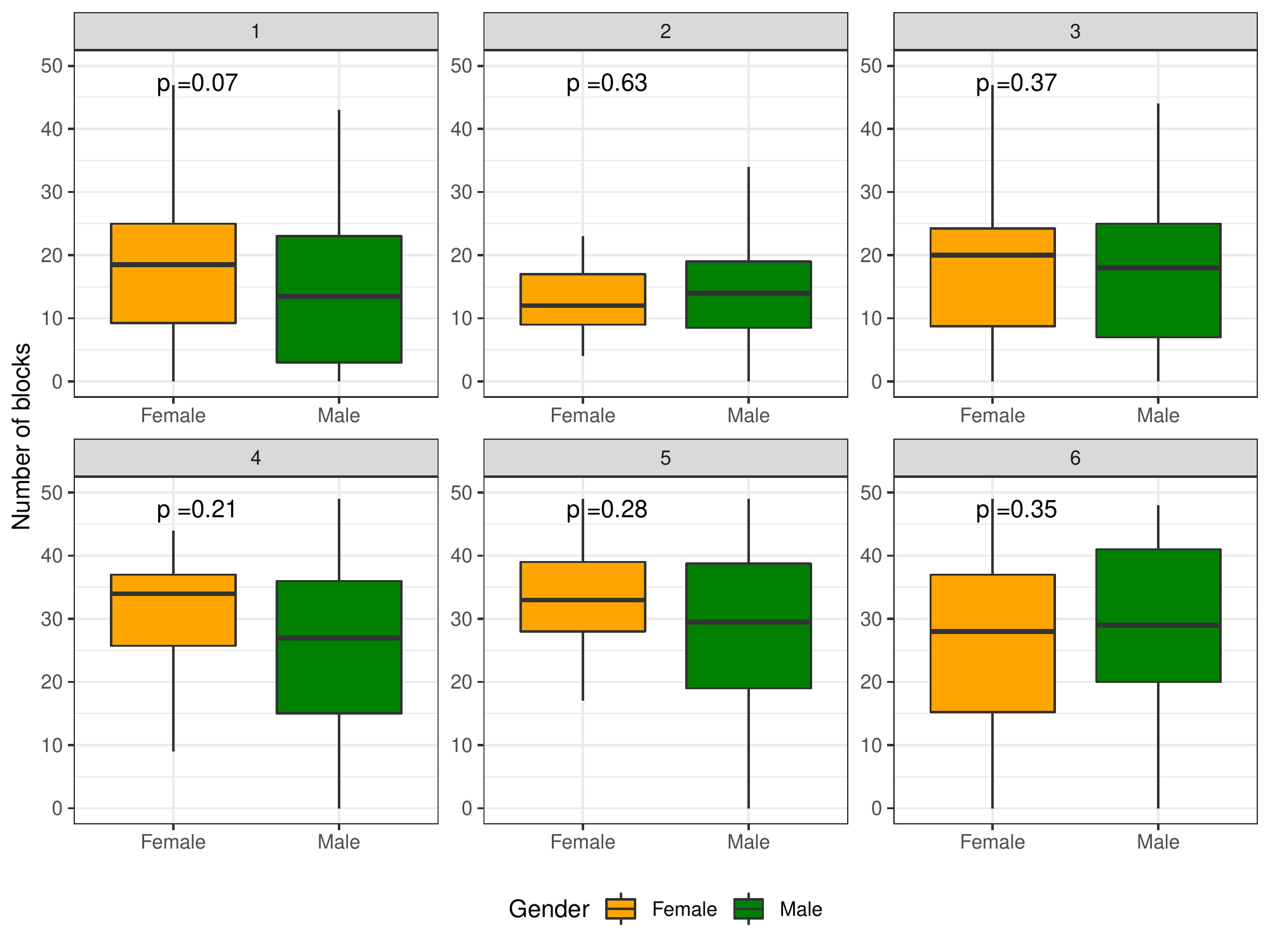}
	\caption{Distribution of the number of blocks among sessions/genders.}
	\label{fig:blocks}
\end{figure}

\begin{figure*}[tb]
	\centering
	\includegraphics[width=0.8\textwidth]{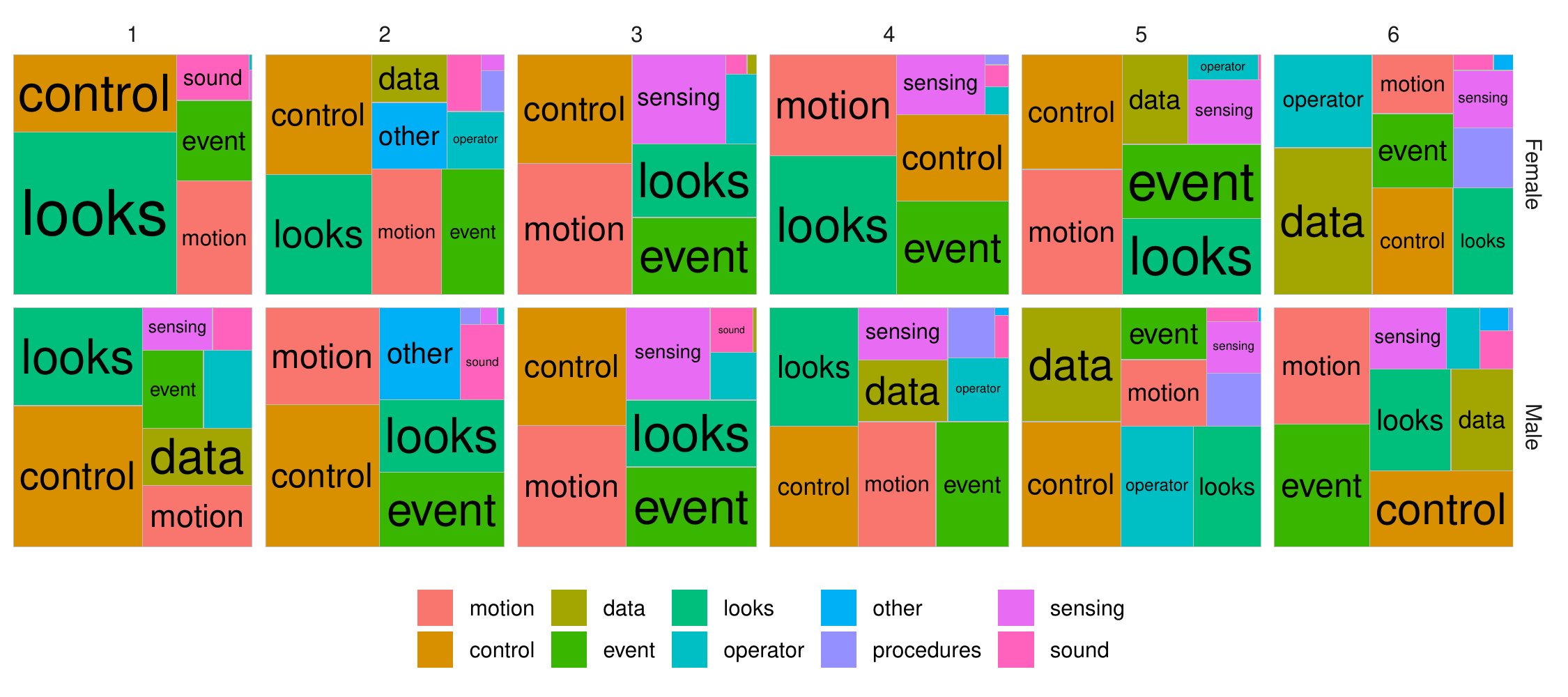}
	\caption{Distribution of block types among session/gender.}
	\label{fig:sessiontypes}
\end{figure*}

\subsubsection{Engagement} 
\label{sec:rq3engagement}
\Cref{fig:addtaskcomparison} summarizes the completion of the additional (optional) tasks throughout the course per gender.
Girls completed the additional task (\cref{sec:contents}) more often than boys throughout the entire course ($p = 0.23$). 
The largest difference between the genders can be seen in the additional task of sessions four and five. 
%
%
Beyond the optional homework tasks, 
35 female students and 30 male students also created projects on their own independently of any tasks set. Especially after the first   (f: 9, m: 7) and the last session (f: 11, m: 8) the creation of self-initiative projects was highest.
%
%

When contrasting the students' self-concept with the actual results, it is
interesting to note that the same number of children of both genders claimed to
engaged \emph{a little bit} with \Scratch and girls indicated less to engage
\emph{often} with it (\cref{engagement}), when the implementation of the
additional tasks and additional projects shows us
(\cref{fig:addtaskcomparison}) that girls actually engaged \emph{more} with
\Scratch using their \Scratch account across the board!


\subsubsection{Code Metrics}

\Cref{fig:blocks} illustrates that the number of blocks used by the students for the defined tasks increases slightly with each task, and is relatively balanced, with girls tending to use slightly more blocks.
Up to the third session, the block types used by both genders
(\cref{fig:sessiontypes}) are very similar. However, for the more complex and
less constrained successive tasks, there are differences: Especially in the
fourth and fifth sessions, girls used many \textit{looks} blocks such as changing the costume of the sprites, which
generally represents the type of block they use preferrably~\cite{robertson2012, funke2017,
grassl2021}. This is particularly interesting since explicit attention was paid
to a gender-neutral course design in terms of topics and project types.
%
The use of \textit{data} blocks is also interesting because the concept of
variables was not introduced until the fifth session, but is applied by both
girls and boys in prior lessons. Girls in particular seem to have internalized
variables in the last lesson, as they use them most often for implementing their
own world.
%
%
%
%
%

\begin{figure}[tb]
\centering
\includegraphics[width=\columnwidth]{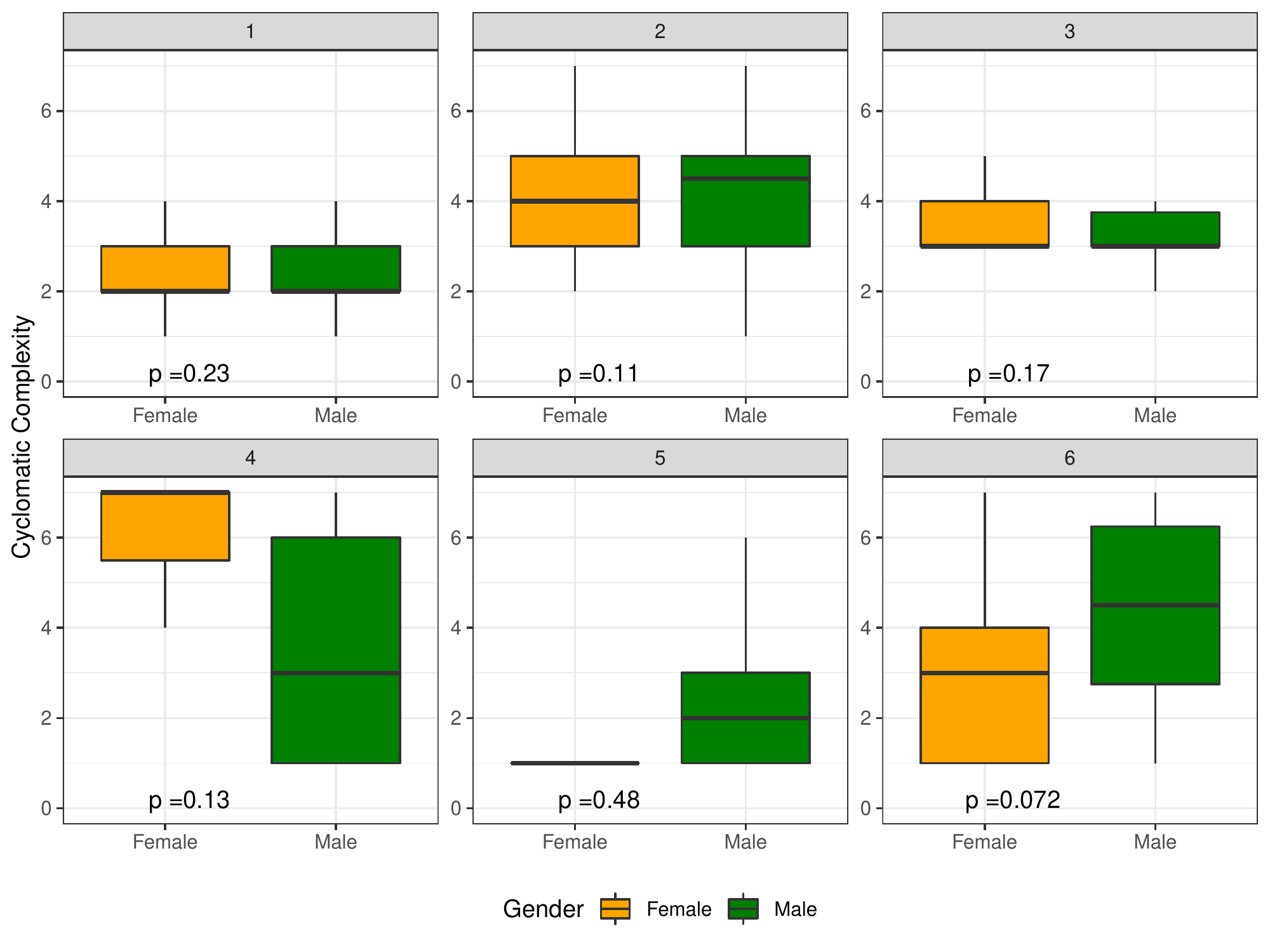}
\caption{Distribution of code complexity measured by ICC.}  
\label{fig:complexity}
\end{figure}

Considering the code complexity of the implemented tasks
(\cref{fig:complexity}), the difference is small for the first three tasks,
which do not allow for much variety. However, the girls' programs are considerably more
complex for the fourth session. This is likely because more than half of these
projects include code resulting from the additional task (\cref{fig:addtaskcomparison}). which involved incorporating different stage changes to represent different levels, and incorporating different event blocks that have different ways of starting a script when switching to a new level.
This also explains the high
proportion of looks, motion, control and event blocks (\cref{fig:sessiontypes}).
However, the girls' projects in the fifth and sixth session show a distinctly
lower complexity compared to the boys' projects. This seems to be because the
girls implemented only the most necessary code elements and then focused more on
adding other elements such as decorations or signs without further
functionality.
%



\rqsummary{RQ3}{Both genders use slightly different types of blocks and have programs of varying complexity. Girls provided correct and additional implementations more often than the boys.}

As girls follow tasks more attentively their extra work should be acknowledged, 
whereas boys should be encouraged to also follow tasks attentatively. When setting open tasks it is important to include playful minimum requirements, as otherwise the girls' knowledge level may stagnate, causing problems with respect to long-term interest.


\subsection{RQ4: Creativity}
\label{sec:rq4}


\begin{figure}[tb]
\centering
\includegraphics[width=\columnwidth]{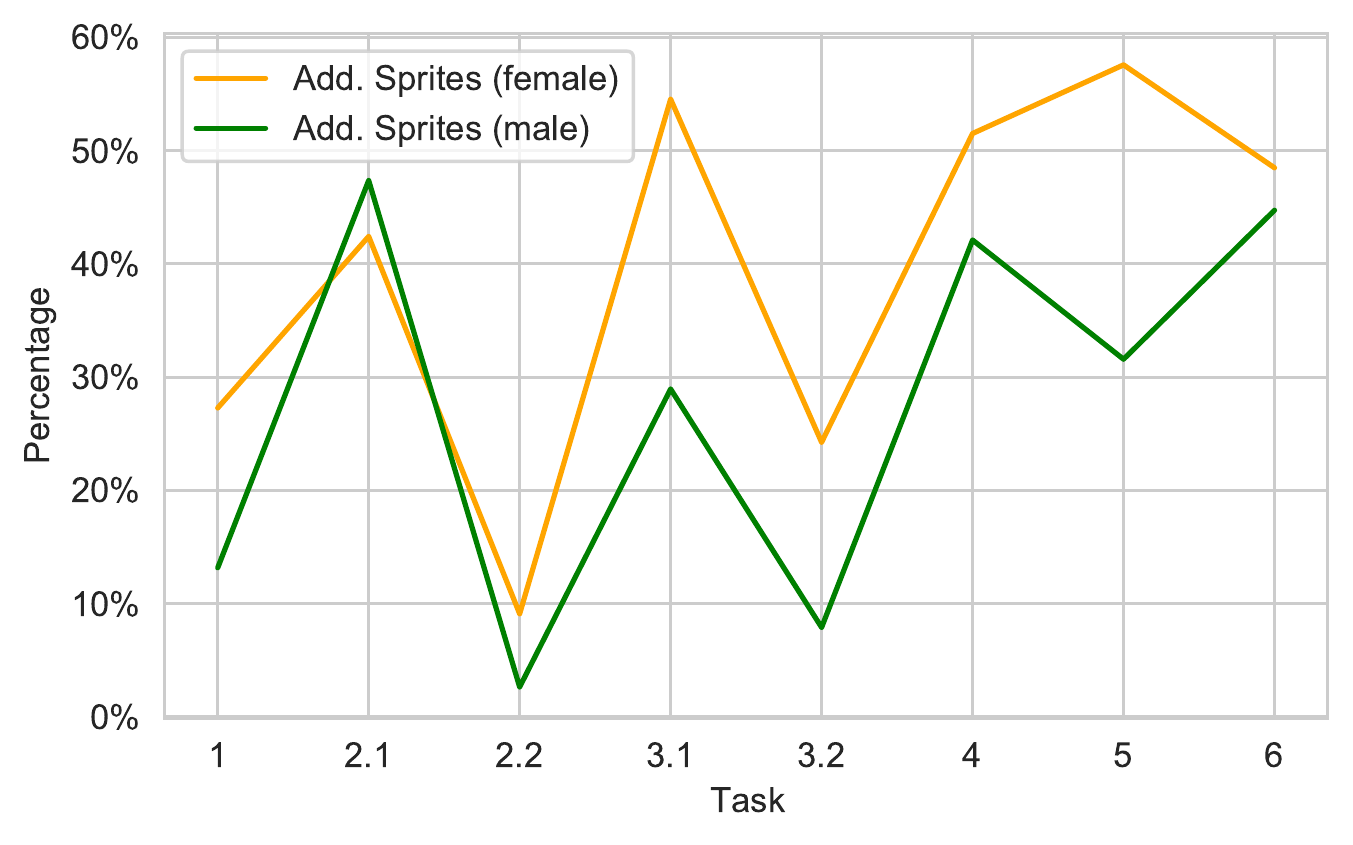}
\caption{Distribution of creating an own sprite.}  
\label{fig:addsprites}
\end{figure}

\begin{figure}[tb]
\centering
\subfloat[\label{sprite4f} Sprite names for session 4 (f).]{\includegraphics[width=0.48\linewidth]{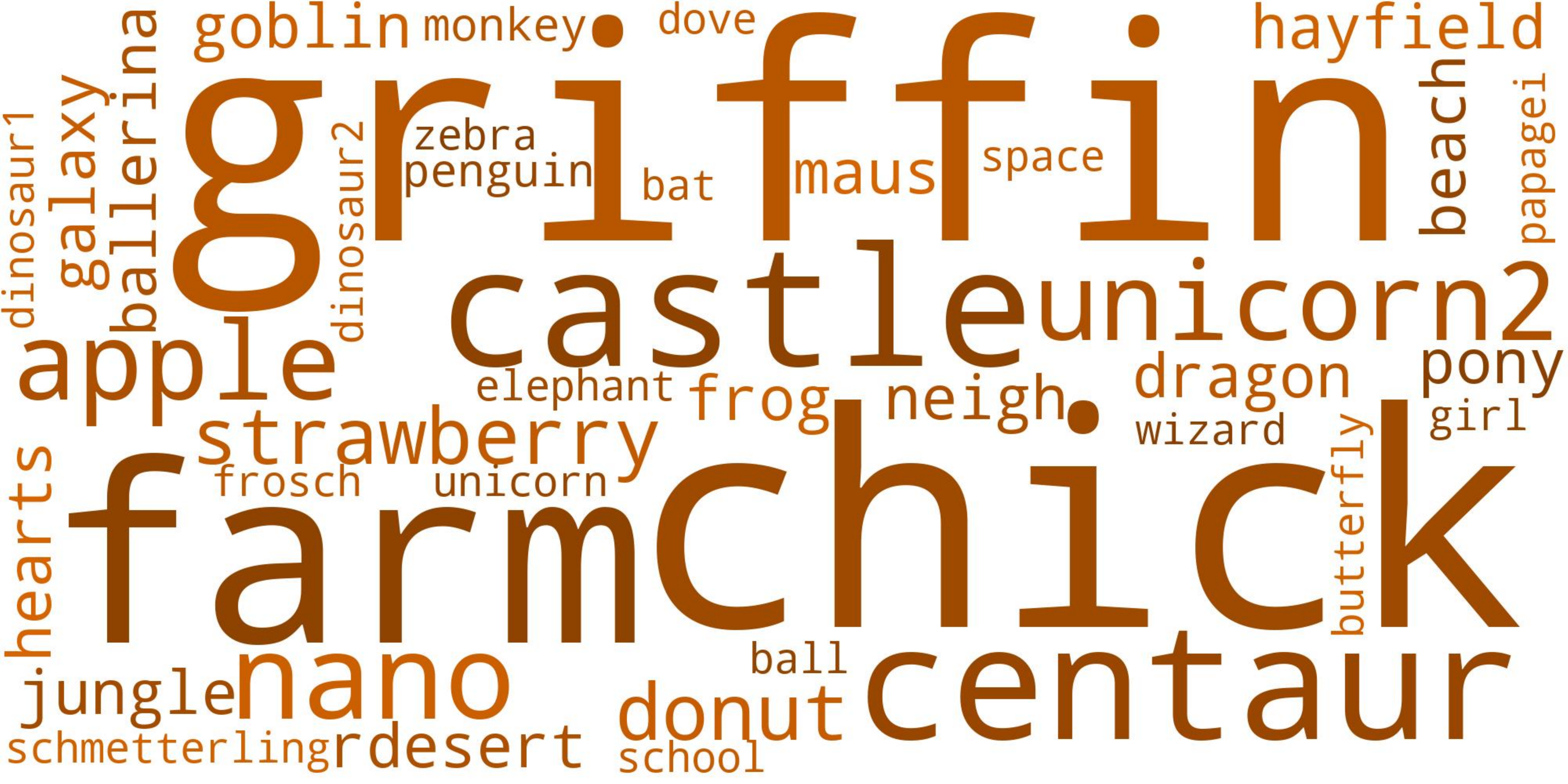}}
\quad
\subfloat[\label{sprite4m} Sprite names for session 4 (m).]{\includegraphics[width=0.48\linewidth]{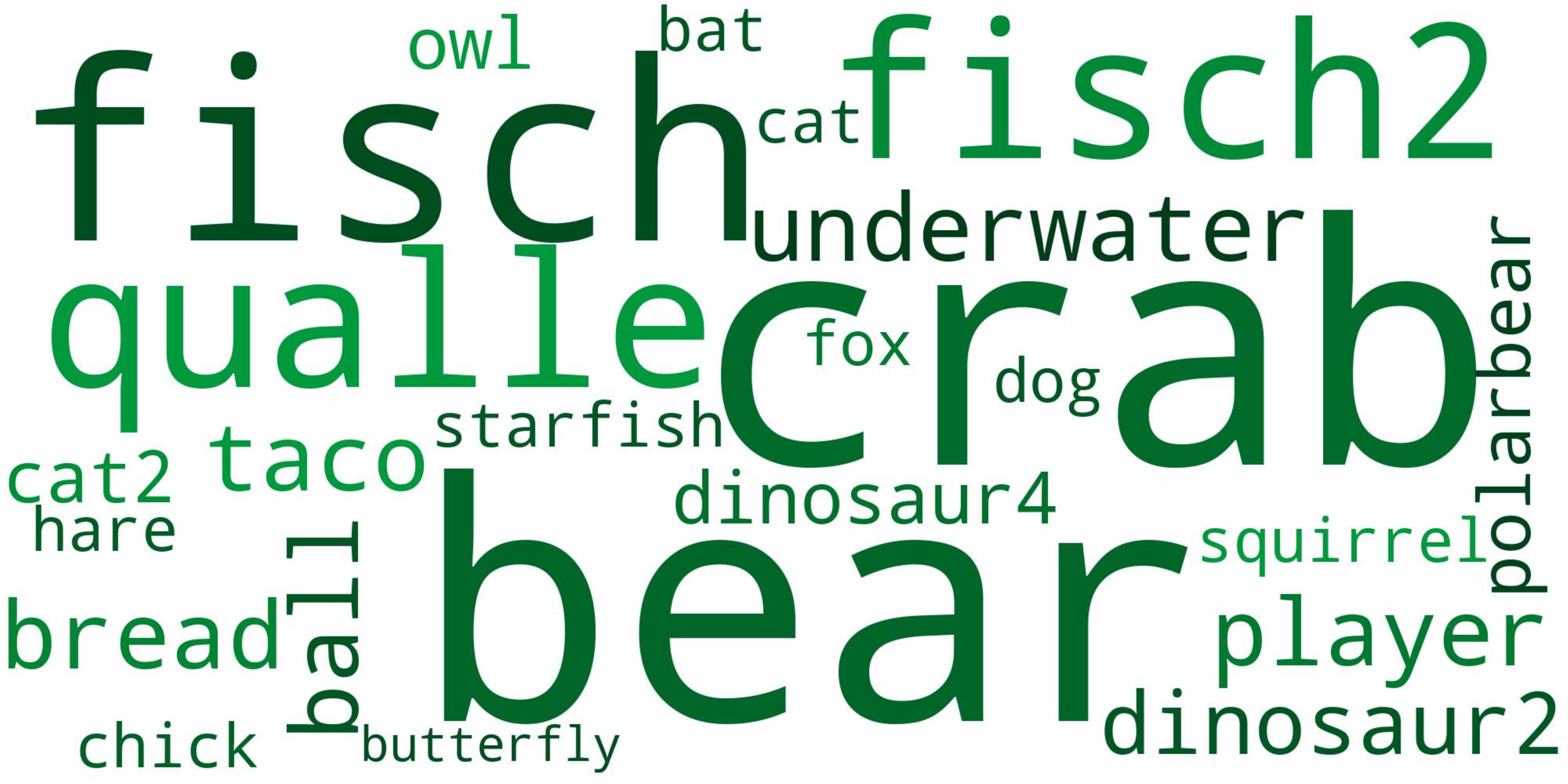}}
\quad
\subfloat[\label{sprite5f} Sprite names for session 5 (f).]{\includegraphics[width=0.48\linewidth]{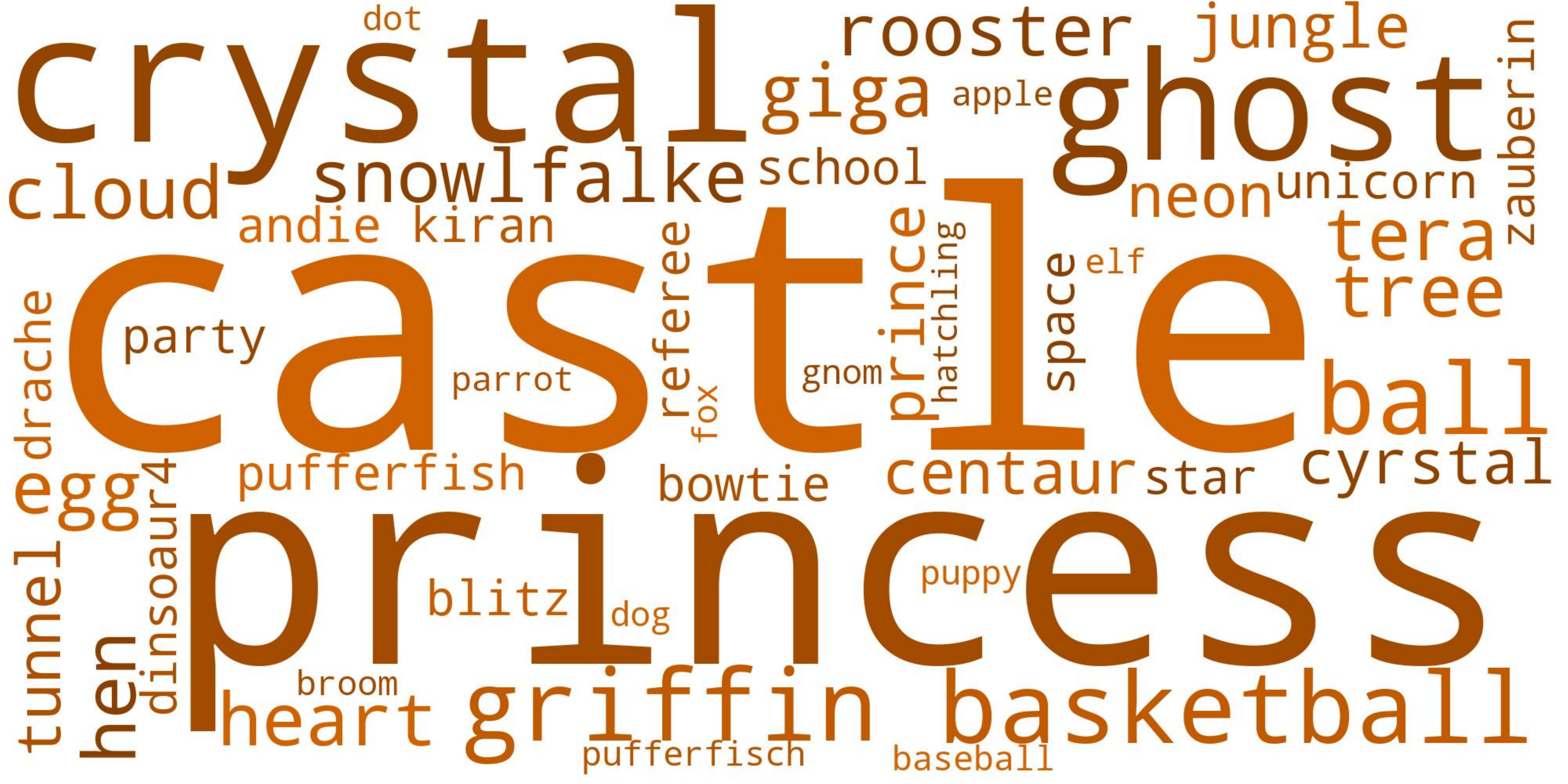}}
\quad
\subfloat[\label{sprite5m} Sprite names for session 5 (m).]{\includegraphics[width=0.48\linewidth]{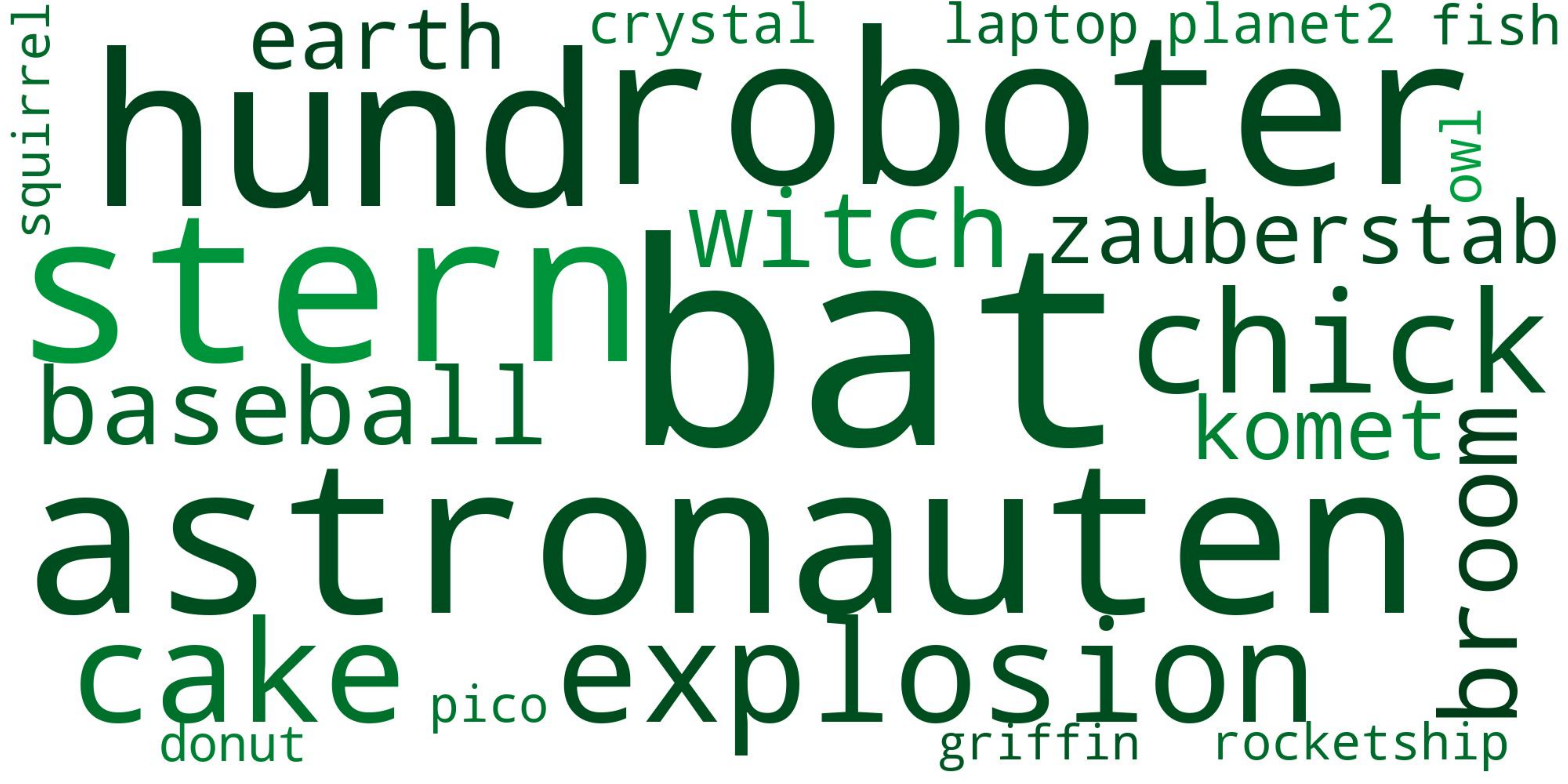}}
\quad
\subfloat[\label{sprite6f} Sprite names for session 6 (f).]{\includegraphics[width=0.48\linewidth]{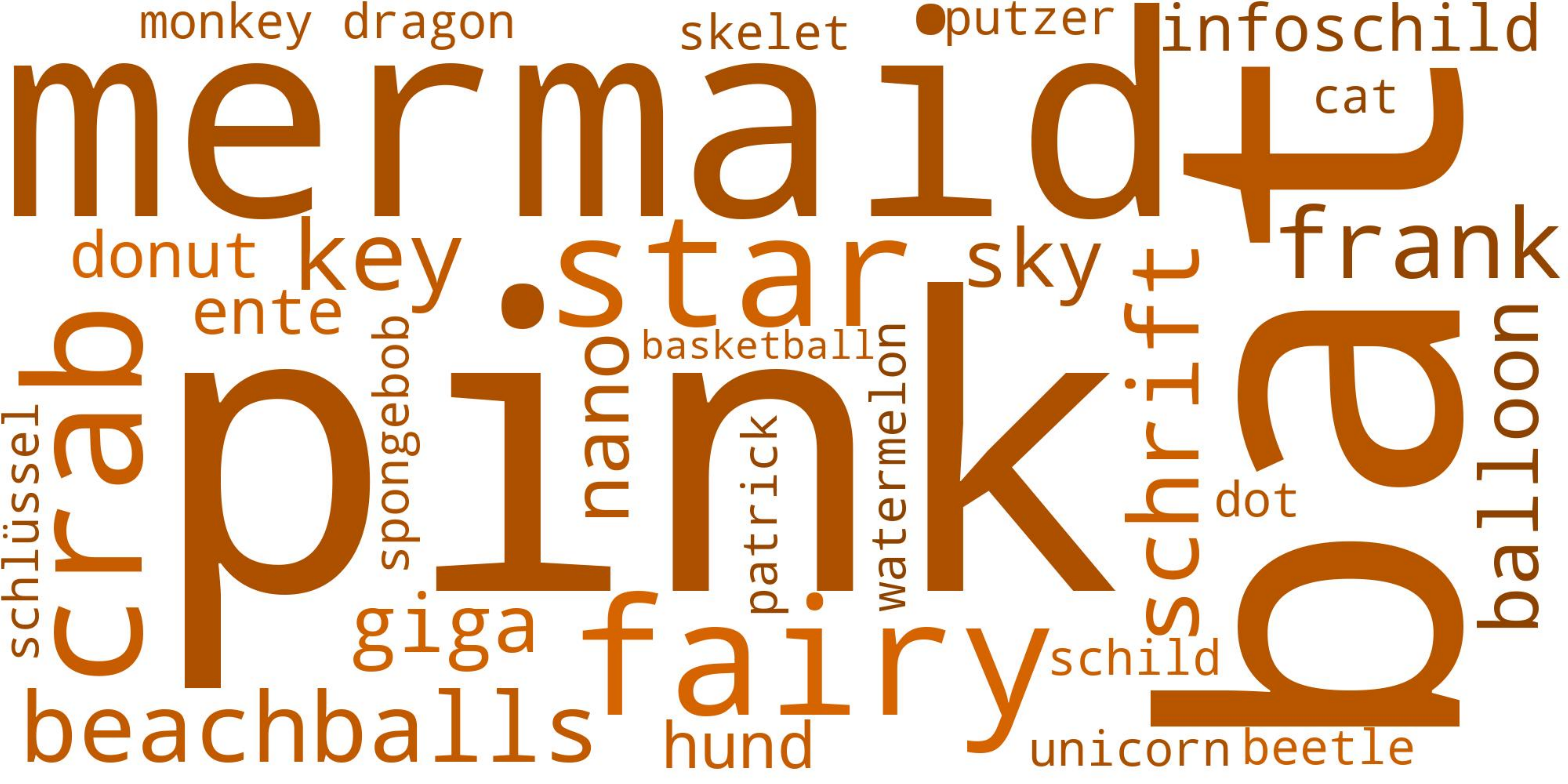}}
\quad
\subfloat[\label{sprite6m} Sprite names for session 6 (m).]{\includegraphics[width=0.48\linewidth]{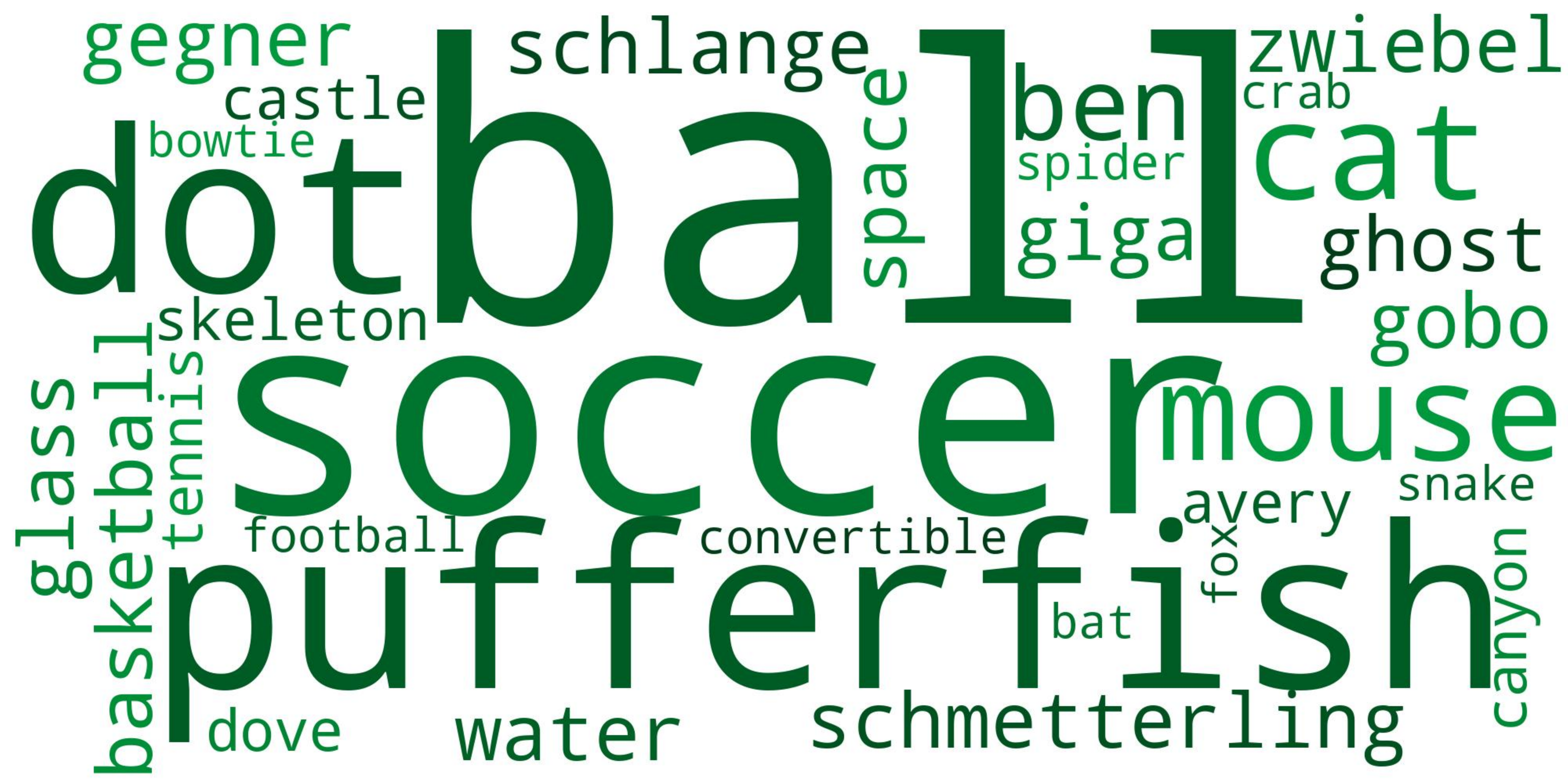}}
\caption{Distribution of the names of the self-chosen sprites, where the size of the name indicates its frequency.}  
\label{fig:wordclouds}
\end{figure}

The sprites and backgrounds used in the programs provide insights into the creative ideas implemented by the children, which is relevant as it is important to allow learners to express their own creativity~\cite{roque2016a, grassl2021} in order to achieve the best possible learning and motivational success.
Girls use more of their own sprites than boys ($p=0.21$) in all but the second session's introduction task (\cref{fig:addsprites}). 
This suggests that the girls have adapted and tweaked the projects
according to their own preferences and therefore have richer
programs~\cite{robertson2012}.
%
%
In particular for the last three sessions, which allowed for more creative freedom, the choice of sprites provide insights into the topics chosen by the students:
The fourth task is a catching game, demonstrated with a cat and mouse. Both genders remained committed to animal sprites and use chicks, penguin, butterfly and monkey (\cref{sprite4f}) as well as bear, crab, fish, owl, fox, polar bear and squirrels (\cref{sprite4m}). Both genders also use fantasy creatures, with girls using unicorns and centaur and boys using dinosaurs.  
 The fifth task is a simple arcade game introduced in two versions, space or greenland scenario. Girls create programs with the setting in a castle featuring princesses, crystals, griffins and ghosts (\cref{sprite5f}) while boys preferred the space theme and added robots, dogs, bats and explosion (\cref{sprite5m}).
In the last session, the children were asked to create their own world. The world of girls seems to be characterized by the color pink as well as fantasy creatures like fairies and mermaids, with bats in it as well (\cref{sprite6f}). The boys use ball as sprite as well as soccer as topic (\cref{sprite6m}). There are many different animals in both worlds.

Clearly, both girls and boys resort to stereotypical topics when given creative
freedom. This is particularly interesting since we placed explicit emphasis on
using gender-neutral topics in the course design and depicting neither implicit
nor explicit female or male sprites or stereotypical topics. 


%

\rqsummary{RQ4}{Girls use their own sprites more than boys. Both genders like to use animals as well as stereotypical preferences.}

Our observations suggest that gender-specific interests are already
internalized at a young age and students express them---analogous to other
media---in programming. It is therefore desirable for teachers to ensure that tasks are
not gender-specific, so that as many students as possible feel addressed and
can apply their creativity without bias.

\section{Conclusions and Future Work}
\label{conclusions}

We designed and implemented an online introductory programming course in
\Scratch with gender-neutral tasks covering the main programming concepts. 
Since both genders equally enjoyed the course and found the
level of difficulty appropriate, this suggests that the course design is
successful even with the challenges of a virtual setting.
%

Our data shows no significant differences in how girls participate in \allfemale vs. \mixed
groups. In line with previous studies, girls were reserved in the first
sessions, but contributed more than boys from halfway through the course---both
in questions and answers. 
%
Similarly, the rate of completed tasks was comparable, although girls slightly
had the edge while boys' projects are more often missing. Furthermore, girls,
regardless of the group constellation, considerably more often implemented the
additional task and also additional projects during the sessions.
%
In terms of creativity, girls used their own sprites more often. Even in a
gender-neutral course setting both genders chose topics in their projects that
reflect stereotypical established norms from society.

An important question for future research is whether the observation that girls contributed more often and performed more additional tasks in our study is influenced more
by the course design or the online format. We are aware of the limited generalizability of our observations, and since the implementation of extra-curricular courses and the participation of sufficient girls is challenging, we encourage replication studies using our replication package online.\footnote{\url{https://github.com/se2p/scratch-online-course}}

In addition, the effects of working in small groups could be investigated with the concept of
pair programming, which has already been shown in some small
studies~\cite{lewis2015, zhong2016, papadakis2018a, ying2019}.


%

Our experience has taught us that we have to reach out not only to the girls, but also to their environment, especially their parents:
Many parents have misconceptions about programming and computer science~\cite{greifenstein2021}, so it could be helpful to convince them as well, as a very motivated student (12f) from the \allfemale group said in the last session about the subsequent advanced course we offered: ``There is a continuation? My father signed me up for the course here because he said I should do that, but my mother said that will definitely not interest me''---but it definitely did.


\vspace{-0.3em}
\begin{acks}
\vspace{-0.3em}We thank Verena Ewald and Florian Oberm\"uller for their contribution to the data collection. 
This work is supported by the Federal Ministry of Education and Research
through project ``primary::programming'' (01JA2021) as
part of the ``Qualitätsoffensive Lehrerbildung'', a joint initiative of the
Federal Government and the Länder. The authors are responsible for the content
of this publication.
\end{acks}

\balance

 \bibliographystyle{ACM-Reference-Format}
 \bibliography{references}


\begin{thebibliography}{45}


\ifx \showCODEN    \undefined \def \showCODEN     #1{\unskip}     \fi
\ifx \showDOI      \undefined \def \showDOI       #1{#1}\fi
\ifx \showISBNx    \undefined \def \showISBNx     #1{\unskip}     \fi
\ifx \showISBNxiii \undefined \def \showISBNxiii  #1{\unskip}     \fi
\ifx \showISSN     \undefined \def \showISSN      #1{\unskip}     \fi
\ifx \showLCCN     \undefined \def \showLCCN      #1{\unskip}     \fi
\ifx \shownote     \undefined \def \shownote      #1{#1}          \fi
\ifx \showarticletitle \undefined \def \showarticletitle #1{#1}   \fi
\ifx \showURL      \undefined \def \showURL       {\relax}        \fi
\providecommand\bibfield[2]{#2}
\providecommand\bibinfo[2]{#2}
\providecommand\natexlab[1]{#1}
\providecommand\showeprint[2][]{arXiv:#2}

\bibitem[\protect\citeauthoryear{Albusays, Bjorn, Dabbish, Ford, {Murphy-Hill},
  Serebrenik, and Storey}{Albusays et~al\mbox{.}}{2021}]%
        {albusays2021}
\bibfield{author}{\bibinfo{person}{K. Albusays}, \bibinfo{person}{P. Bjorn},
  \bibinfo{person}{L. Dabbish}, \bibinfo{person}{D. Ford}, \bibinfo{person}{E.
  {Murphy-Hill}}, \bibinfo{person}{A. Serebrenik}, {and} \bibinfo{person}{M.-A.
  Storey}.} \bibinfo{year}{2021}\natexlab{}.
\newblock \showarticletitle{The {{Diversity Crisis}} in {{Software
  Development}}}.
\newblock \bibinfo{journal}{\emph{IEEE Software}} \bibinfo{volume}{38},
  \bibinfo{number}{2} (\bibinfo{date}{March} \bibinfo{year}{2021}),
  \bibinfo{pages}{19--25}.
\newblock
\showISSN{1937-4194}


\bibitem[\protect\citeauthoryear{Bandura}{Bandura}{1977}]%
        {bandura1977}
\bibfield{author}{\bibinfo{person}{A. Bandura}.}
  \bibinfo{year}{1977}\natexlab{}.
\newblock \showarticletitle{Self-Efficacy: Toward a {{Unifying Theory}} of
  {{Behavioral Change}}}.
\newblock \bibinfo{journal}{\emph{Psychological Review}}  \bibinfo{volume}{84}
  (\bibinfo{year}{1977}), \bibinfo{pages}{191--215}.
\newblock


\bibitem[\protect\citeauthoryear{Beckwith, Burnett, Wiedenbeck, Cook, Sorte,
  and Hastings}{Beckwith et~al\mbox{.}}{2005}]%
        {beckwith2005}
\bibfield{author}{\bibinfo{person}{L. Beckwith}, \bibinfo{person}{M. Burnett},
  \bibinfo{person}{S. Wiedenbeck}, \bibinfo{person}{C. Cook},
  \bibinfo{person}{S. Sorte}, {and} \bibinfo{person}{M. Hastings}.}
  \bibinfo{year}{2005}\natexlab{}.
\newblock \showarticletitle{Effectiveness of end-user debugging software
  features: Are there gender issues?}. In \bibinfo{booktitle}{\emph{Proceedings
  of the SIGCHI Conference on Human Factors in Computing Systems}}.
  \bibinfo{pages}{869--878}.
\newblock


\bibitem[\protect\citeauthoryear{Betz and Hackett}{Betz and Hackett}{1986}]%
        {betz1986}
\bibfield{author}{\bibinfo{person}{N.~E. Betz} {and} \bibinfo{person}{G.
  Hackett}.} \bibinfo{year}{1986}\natexlab{}.
\newblock \showarticletitle{Applications of self-efficacy theory to
  understanding career choice behavior}.
\newblock \bibinfo{journal}{\emph{Journal of social and clinical psychology}}
  \bibinfo{volume}{4}, \bibinfo{number}{3} (\bibinfo{year}{1986}),
  \bibinfo{pages}{279--289}.
\newblock


\bibitem[\protect\citeauthoryear{Beyer}{Beyer}{2014}]%
        {beyer2014}
\bibfield{author}{\bibinfo{person}{S. Beyer}.} \bibinfo{year}{2014}\natexlab{}.
\newblock \showarticletitle{Why Are Women Underrepresented in {{Computer
  Science}}? Gender Differences in Stereotypes, Self-Efficacy, Values, and
  Interests and Predictors of Future {{CS}} Course-Taking and Grades}.
\newblock \bibinfo{journal}{\emph{Computer Science Education}}
  \bibinfo{volume}{24}, \bibinfo{number}{2-3} (\bibinfo{date}{July}
  \bibinfo{year}{2014}), \bibinfo{pages}{153--192}.
\newblock
\showISSN{0899-3408, 1744-5175}


\bibitem[\protect\citeauthoryear{Bosu and Sultana}{Bosu and Sultana}{2019}]%
        {bosu2019}
\bibfield{author}{\bibinfo{person}{A. Bosu} {and} \bibinfo{person}{K.~Z.
  Sultana}.} \bibinfo{year}{2019}\natexlab{}.
\newblock \showarticletitle{Diversity and {{Inclusion}} in {{Open Source
  Software}} ({{OSS}}) {{Projects}}: Where {{Do We Stand}}?}. In
  \bibinfo{booktitle}{\emph{2019 {{ACM}}/{{IEEE International Symposium}} on
  {{Empirical Software Engineering}} and {{Measurement}} ({{ESEM}})}}.
  \bibinfo{pages}{1--11}.
\newblock
\showISSN{1949-3789}


\bibitem[\protect\citeauthoryear{{\c{C}}ak{\i}r, Gass, Foster, and
  Lee}{{\c{C}}ak{\i}r et~al\mbox{.}}{2017}]%
        {ccakir2017}
\bibfield{author}{\bibinfo{person}{N.~A. {\c{C}}ak{\i}r}, \bibinfo{person}{A.
  Gass}, \bibinfo{person}{A. Foster}, {and} \bibinfo{person}{F.~J. Lee}.}
  \bibinfo{year}{2017}\natexlab{}.
\newblock \showarticletitle{Development of a game-design workshop to promote
  young girls' interest towards computing through identity exploration}.
\newblock \bibinfo{journal}{\emph{Computers \& Education}}
  \bibinfo{volume}{108} (\bibinfo{year}{2017}), \bibinfo{pages}{115--130}.
\newblock


\bibitem[\protect\citeauthoryear{Cen, Ruta, Powell, and Ng}{Cen
  et~al\mbox{.}}{2014}]%
        {cen2014}
\bibfield{author}{\bibinfo{person}{L. Cen}, \bibinfo{person}{D. Ruta},
  \bibinfo{person}{L. Powell}, {and} \bibinfo{person}{J. Ng}.}
  \bibinfo{year}{2014}\natexlab{}.
\newblock \showarticletitle{Does Gender Matter for Collaborative Learning?}. In
  \bibinfo{booktitle}{\emph{2014 {{IEEE International Conference}} on
  {{Teaching}}, {{Assessment}} and {{Learning}} for {{Engineering}}
  ({{TALE}})}}. \bibinfo{pages}{433--440}.
\newblock


\bibitem[\protect\citeauthoryear{Crombie, Abarbanel, and Anderson}{Crombie
  et~al\mbox{.}}{2000}]%
        {crombie2000}
\bibfield{author}{\bibinfo{person}{G. Crombie}, \bibinfo{person}{T. Abarbanel},
  {and} \bibinfo{person}{C. Anderson}.} \bibinfo{year}{2000}\natexlab{}.
\newblock \showarticletitle{All-{{Female Computer Science}}}.
\newblock \bibinfo{journal}{\emph{Science Teacher}} \bibinfo{volume}{67},
  \bibinfo{number}{3} (\bibinfo{year}{2000}), \bibinfo{pages}{40--43}.
\newblock
\showISSN{0036-8555}


\bibitem[\protect\citeauthoryear{Fields, Kafai, and Giang}{Fields
  et~al\mbox{.}}{[n.d.]}]%
        {fields}
\bibfield{author}{\bibinfo{person}{D.~A. Fields}, \bibinfo{person}{Y.~B.
  Kafai}, {and} \bibinfo{person}{M.~T. Giang}.}
  \bibinfo{year}{[n.d.]}\natexlab{}.
\newblock \showarticletitle{Youth {{Computational Participation}} in the
  {{Wild}}: Understanding {{Experience}} and {{Equity}} in {{Participating}}
  and {{Programming}} in the {{Online Scratch Community}}}.
\newblock \bibinfo{journal}{\emph{ACM Transactions on Computing Education}}
  \bibinfo{volume}{17}, \bibinfo{number}{3} (\bibinfo{year}{[n.\,d.]}),
  \bibinfo{pages}{22}.
\newblock


\bibitem[\protect\citeauthoryear{Fields, Kafai, Strommer, Wolf, and
  Seiner}{Fields et~al\mbox{.}}{2014}]%
        {fields2014}
\bibfield{author}{\bibinfo{person}{D.~A. Fields}, \bibinfo{person}{Y.~B.
  Kafai}, \bibinfo{person}{A. Strommer}, \bibinfo{person}{E. Wolf}, {and}
  \bibinfo{person}{B. Seiner}.} \bibinfo{year}{2014}\natexlab{}.
\newblock \showarticletitle{{{INTERACTIVE STORYTELLING FOR PROMOTING CREATIVE
  EXPRESSION IN MEDIA AND CODING IN YOUTH ONLINE COLLABORATIVES IN SCRATCH}}}.
\newblock  (\bibinfo{year}{2014}), \bibinfo{pages}{11}.
\newblock


\bibitem[\protect\citeauthoryear{Fraser, Heuer, K{\"o}rber, Wasmeier,
  et~al\mbox{.}}{Fraser et~al\mbox{.}}{2021}]%
        {fraser2021litterbox}
\bibfield{author}{\bibinfo{person}{G. Fraser}, \bibinfo{person}{U. Heuer},
  \bibinfo{person}{N. K{\"o}rber}, \bibinfo{person}{E. Wasmeier},
  {et~al\mbox{.}}} \bibinfo{year}{2021}\natexlab{}.
\newblock \showarticletitle{LitterBox: A Linter for Scratch Programs}. In
  \bibinfo{booktitle}{\emph{2021 IEEE/ACM 43rd International Conference on
  Software Engineering: Software Engineering Education and Training
  (ICSE-SEET)}}. IEEE, \bibinfo{pages}{183--188}.
\newblock


\bibitem[\protect\citeauthoryear{Funke and Geldreich}{Funke and
  Geldreich}{2017}]%
        {funke2017}
\bibfield{author}{\bibinfo{person}{A. Funke} {and} \bibinfo{person}{K.
  Geldreich}.} \bibinfo{year}{2017}\natexlab{}.
\newblock \showarticletitle{Gender {{Differences}} in {{Scratch Programs}} of
  {{Primary School Children}}}. In \bibinfo{booktitle}{\emph{Proceedings of the
  12th {{Workshop}} on {{Primary}} and {{Secondary Computing Education}}}}.
  \bibinfo{publisher}{{ACM}}, \bibinfo{address}{{Nijmegen Netherlands}},
  \bibinfo{pages}{57--64}.
\newblock
\showISBNx{978-1-4503-5428-8}


\bibitem[\protect\citeauthoryear{Gonz{\'a}lez-G{\'o}mez, Guardiola,
  Rodr{\'\i}guez, and Alonso}{Gonz{\'a}lez-G{\'o}mez et~al\mbox{.}}{2012}]%
        {gonzalez2012}
\bibfield{author}{\bibinfo{person}{F. Gonz{\'a}lez-G{\'o}mez},
  \bibinfo{person}{J. Guardiola}, \bibinfo{person}{{\'O}.~M. Rodr{\'\i}guez},
  {and} \bibinfo{person}{M.~{\'A}.~M. Alonso}.}
  \bibinfo{year}{2012}\natexlab{}.
\newblock \showarticletitle{Gender differences in e-learning satisfaction}.
\newblock \bibinfo{journal}{\emph{Computers \& Education}}
  \bibinfo{volume}{58}, \bibinfo{number}{1} (\bibinfo{year}{2012}),
  \bibinfo{pages}{283--290}.
\newblock


\bibitem[\protect\citeauthoryear{Gra{\ss}l, Geldreich, and Fraser}{Gra{\ss}l
  et~al\mbox{.}}{2021}]%
        {grassl2021}
\bibfield{author}{\bibinfo{person}{I. Gra{\ss}l}, \bibinfo{person}{K.
  Geldreich}, {and} \bibinfo{person}{G. Fraser}.}
  \bibinfo{year}{2021}\natexlab{}.
\newblock \showarticletitle{Data-driven Analysis of Gender Differences and
  Similarities in Scratch Programs}. In \bibinfo{booktitle}{\emph{Proceedings
  of the 16th Workshop in Primary and Secondary Computing Education (WiPSCE)}}.
  \bibinfo{publisher}{ACM}.
\newblock
\newblock
\shownote{To appear.}


\bibitem[\protect\citeauthoryear{Greifenstein, Gra{\ss}l, and
  Fraser}{Greifenstein et~al\mbox{.}}{2021}]%
        {greifenstein2021}
\bibfield{author}{\bibinfo{person}{L. Greifenstein}, \bibinfo{person}{I.
  Gra{\ss}l}, {and} \bibinfo{person}{G. Fraser}.}
  \bibinfo{year}{2021}\natexlab{}.
\newblock \showarticletitle{Challenging but Full of Opportunities: Teachers’
  Perspectives on Programming in Primary Schools}. In
  \bibinfo{booktitle}{\emph{21st Koli Calling International Conference on
  Computing Education Research}}. \bibinfo{pages}{1--10}.
\newblock


\bibitem[\protect\citeauthoryear{Grover, Pea, and Cooper}{Grover
  et~al\mbox{.}}{2015}]%
        {grover2015}
\bibfield{author}{\bibinfo{person}{S. Grover}, \bibinfo{person}{R. Pea}, {and}
  \bibinfo{person}{S. Cooper}.} \bibinfo{year}{2015}\natexlab{}.
\newblock \showarticletitle{Designing for deeper learning in a blended computer
  science course for middle school students}.
\newblock \bibinfo{journal}{\emph{Computer science education}}
  \bibinfo{volume}{25}, \bibinfo{number}{2} (\bibinfo{year}{2015}),
  \bibinfo{pages}{199--237}.
\newblock


\bibitem[\protect\citeauthoryear{Hermans and Aivaloglou}{Hermans and
  Aivaloglou}{2017}]%
        {hermans2017b}
\bibfield{author}{\bibinfo{person}{F. Hermans} {and} \bibinfo{person}{E.
  Aivaloglou}.} \bibinfo{year}{2017}\natexlab{}.
\newblock \showarticletitle{Teaching {{Software Engineering Principles}} to
  {{K}}-12 {{Students}}: A {{MOOC}} on {{Scratch}}}. In
  \bibinfo{booktitle}{\emph{2017 {{IEEE}}/{{ACM}} 39th {{International
  Conference}} on {{Software Engineering}}: Software {{Engineering Education}}
  and {{Training Track}} ({{ICSE}}-{{SEET}})}}. \bibinfo{pages}{13--22}.
\newblock


\bibitem[\protect\citeauthoryear{Hubwieser, Hubwieser, and Graswald}{Hubwieser
  et~al\mbox{.}}{2016}]%
        {hubwieser2016}
\bibfield{author}{\bibinfo{person}{P. Hubwieser}, \bibinfo{person}{E.
  Hubwieser}, {and} \bibinfo{person}{D. Graswald}.}
  \bibinfo{year}{2016}\natexlab{}.
\newblock \showarticletitle{How to {{Attract}} the {{Girls}}: Gender-{{Specific
  Performance}} and {{Motivation}} in the {{Bebras Challenge}}}. In
  \bibinfo{booktitle}{\emph{Informatics in {{Schools}}: Improvement of
  {{Informatics Knowledge}} and {{Perception}}}}
  \emph{(\bibinfo{series}{Lecture {{Notes}} in {{Computer Science}}})},
  \bibfield{editor}{\bibinfo{person}{A.~Brodnik} {and}
  \bibinfo{person}{F.~Tort}} (Eds.). \bibinfo{publisher}{{Springer
  International Publishing}}, \bibinfo{address}{{Cham}},
  \bibinfo{pages}{40--52}.
\newblock
\showISBNx{978-3-319-46747-4}


\bibitem[\protect\citeauthoryear{Jones and Clarke}{Jones and Clarke}{1995}]%
        {jones1995}
\bibfield{author}{\bibinfo{person}{T. Jones} {and} \bibinfo{person}{V.~A.
  Clarke}.} \bibinfo{year}{1995}\natexlab{}.
\newblock \showarticletitle{Diversity as a {{Determinant}} of {{Attitudes}}: A
  {{Possible Explanation}} of the {{Apparent Advantage}} of {{Single}}-{{Sex
  Settings}}}.
\newblock \bibinfo{journal}{\emph{Journal of Educational Computing Research}}
  \bibinfo{volume}{12}, \bibinfo{number}{1} (\bibinfo{date}{Jan.}
  \bibinfo{year}{1995}), \bibinfo{pages}{51--64}.
\newblock
\showISSN{0735-6331}


\bibitem[\protect\citeauthoryear{Lewis and Shah}{Lewis and Shah}{2015}]%
        {lewis2015}
\bibfield{author}{\bibinfo{person}{C.~M. Lewis} {and} \bibinfo{person}{N.
  Shah}.} \bibinfo{year}{2015}\natexlab{}.
\newblock \showarticletitle{How {{Equity}} and {{Inequity Can Emerge}} in
  {{Pair Programming}}}. In \bibinfo{booktitle}{\emph{Proceedings of the
  Eleventh Annual {{International Conference}} on {{International Computing
  Education Research}}}}. \bibinfo{publisher}{{ACM}}, \bibinfo{address}{{Omaha
  Nebraska USA}}, \bibinfo{pages}{41--50}.
\newblock
\showISBNx{978-1-4503-3630-7}


\bibitem[\protect\citeauthoryear{Lindberg, Hyde, Petersen, and Linn}{Lindberg
  et~al\mbox{.}}{2010}]%
        {lindberg2010}
\bibfield{author}{\bibinfo{person}{S.~M. Lindberg}, \bibinfo{person}{J.~S.
  Hyde}, \bibinfo{person}{J.~L. Petersen}, {and} \bibinfo{person}{M.~C. Linn}.}
  \bibinfo{year}{2010}\natexlab{}.
\newblock \showarticletitle{New {{Trends}} in {{Gender}} and {{Mathematics
  Performance}}: A {{Meta}}-{{Analysis}}}.
\newblock \bibinfo{journal}{\emph{Psychological bulletin}}
  \bibinfo{volume}{136}, \bibinfo{number}{6} (\bibinfo{date}{Nov.}
  \bibinfo{year}{2010}), \bibinfo{pages}{1123--1135}.
\newblock
\showISSN{0033-2909}


\bibitem[\protect\citeauthoryear{Lishinski, Yadav, Good, and Enbody}{Lishinski
  et~al\mbox{.}}{2016}]%
        {lishinski2016}
\bibfield{author}{\bibinfo{person}{A. Lishinski}, \bibinfo{person}{A. Yadav},
  \bibinfo{person}{J. Good}, {and} \bibinfo{person}{R. Enbody}.}
  \bibinfo{year}{2016}\natexlab{}.
\newblock \showarticletitle{Learning to {{Program}}: Gender {{Differences}} and
  {{Interactive Effects}} of {{Students}}' {{Motivation}}, {{Goals}}, and
  {{Self}}-{{Efficacy}} on {{Performance}}}. In
  \bibinfo{booktitle}{\emph{Proceedings of the 2016 {{ACM Conference}} on
  {{International Computing Education Research}}}}. \bibinfo{publisher}{{ACM}},
  \bibinfo{address}{{Melbourne VIC Australia}}, \bibinfo{pages}{211--220}.
\newblock
\showISBNx{978-1-4503-4449-4}


\bibitem[\protect\citeauthoryear{Margolis and Fisher}{Margolis and
  Fisher}{2002}]%
        {margolis2002}
\bibfield{author}{\bibinfo{person}{J. Margolis} {and} \bibinfo{person}{A.
  Fisher}.} \bibinfo{year}{2002}\natexlab{}.
\newblock \bibinfo{booktitle}{\emph{Unlocking the {{Clubhouse}}: Women in
  {{Computing}}}}.
\newblock \bibinfo{publisher}{{MIT Press}}.
\newblock
\showISBNx{978-0-262-63269-0}


\bibitem[\protect\citeauthoryear{McBroom, Koprinska, and Yacef}{McBroom
  et~al\mbox{.}}{2020}]%
        {mcbroom2020}
\bibfield{author}{\bibinfo{person}{J. McBroom}, \bibinfo{person}{I. Koprinska},
  {and} \bibinfo{person}{K. Yacef}.} \bibinfo{year}{2020}\natexlab{}.
\newblock \showarticletitle{Understanding {{Gender Differences}} to {{Improve
  Equity}} in {{Computer Programming Education}}}. In
  \bibinfo{booktitle}{\emph{Proceedings of the {{Twenty}}-{{Second Australasian
  Computing Education Conference}}}}. \bibinfo{publisher}{{ACM}},
  \bibinfo{address}{{Melbourne VIC Australia}}, \bibinfo{pages}{185--194}.
\newblock
\showISBNx{978-1-4503-7686-0}


\bibitem[\protect\citeauthoryear{McDowell, Werner, Bullock, and
  Fernald}{McDowell et~al\mbox{.}}{2006}]%
        {mcdowell2006}
\bibfield{author}{\bibinfo{person}{C. McDowell}, \bibinfo{person}{L. Werner},
  \bibinfo{person}{H.~E. Bullock}, {and} \bibinfo{person}{J. Fernald}.}
  \bibinfo{year}{2006}\natexlab{}.
\newblock \showarticletitle{Pair programming improves student retention,
  confidence, and program quality}.
\newblock \bibinfo{journal}{\emph{Commun. ACM}} \bibinfo{volume}{49},
  \bibinfo{number}{8} (\bibinfo{year}{2006}), \bibinfo{pages}{90--95}.
\newblock


\bibitem[\protect\citeauthoryear{Mitchell}{Mitchell}{1993}]%
        {mitchell1993}
\bibfield{author}{\bibinfo{person}{M. Mitchell}.}
  \bibinfo{year}{1993}\natexlab{}.
\newblock \showarticletitle{Situational interest: Its multifaceted structure in
  the secondary school mathematics classroom.}
\newblock \bibinfo{journal}{\emph{Journal of educational psychology}}
  \bibinfo{volume}{85}, \bibinfo{number}{3} (\bibinfo{year}{1993}),
  \bibinfo{pages}{424}.
\newblock


\bibitem[\protect\citeauthoryear{Moorman and Johnson}{Moorman and
  Johnson}{2003}]%
        {moorman2003}
\bibfield{author}{\bibinfo{person}{P. Moorman} {and} \bibinfo{person}{E.
  Johnson}.} \bibinfo{year}{2003}\natexlab{}.
\newblock \showarticletitle{Still {{A Stranger Here}}: Attitudes {{Among
  Secondary School Students Towards Computer Science}}}.
\newblock \bibinfo{journal}{\emph{ACM SIGCSE Bulletin}} \bibinfo{volume}{35},
  \bibinfo{number}{3} (\bibinfo{year}{2003}), \bibinfo{pages}{193--197}.
\newblock


\bibitem[\protect\citeauthoryear{Papadakis}{Papadakis}{2018}]%
        {papadakis2018a}
\bibfield{author}{\bibinfo{person}{S. Papadakis}.}
  \bibinfo{year}{2018}\natexlab{}.
\newblock \showarticletitle{Is {{Pair Programming More Effective}} than {{Solo
  Programming}} for {{Secondary Education Novice Programmers}}?: A {{Case
  Study}}}.
\newblock \bibinfo{journal}{\emph{International Journal of Web-Based Learning
  and Teaching Technologies (IJWLTT)}} \bibinfo{volume}{13},
  \bibinfo{number}{1} (\bibinfo{date}{Jan.} \bibinfo{year}{2018}),
  \bibinfo{pages}{1--16}.
\newblock
\showISSN{1548-1093}


\bibitem[\protect\citeauthoryear{REDINE}{REDINE}{2019}]%
        {redine2019}
\bibfield{author}{\bibinfo{person}{REDINE}.} \bibinfo{year}{2019}\natexlab{}.
\newblock \bibinfo{booktitle}{\emph{{Conference Proceedings EDUNOVATIC 2018:
  3rd Virtual International Conference on Education, Innovation and ICT}}}.
\newblock \bibinfo{publisher}{{Adaya Press}}.
\newblock
\showISBNx{978-94-92805-08-9}


\bibitem[\protect\citeauthoryear{Resnick, Maloney, {Monroy-Hern{\'a}ndez},
  Rusk, Eastmond, Brennan, Millner, Rosenbaum, Silver, Silverman, and
  Kafai}{Resnick et~al\mbox{.}}{2009}]%
        {resnick2009a}
\bibfield{author}{\bibinfo{person}{M. Resnick}, \bibinfo{person}{J. Maloney},
  \bibinfo{person}{A. {Monroy-Hern{\'a}ndez}}, \bibinfo{person}{N. Rusk},
  \bibinfo{person}{E. Eastmond}, \bibinfo{person}{K. Brennan},
  \bibinfo{person}{A. Millner}, \bibinfo{person}{E. Rosenbaum},
  \bibinfo{person}{J. Silver}, \bibinfo{person}{B. Silverman}, {and}
  \bibinfo{person}{Y. Kafai}.} \bibinfo{year}{2009}\natexlab{}.
\newblock \showarticletitle{Scratch: Programming for All}.
\newblock \bibinfo{journal}{\emph{Commun. ACM}} \bibinfo{volume}{52},
  \bibinfo{number}{11} (\bibinfo{date}{Nov.} \bibinfo{year}{2009}),
  \bibinfo{pages}{60--67}.
\newblock
\showISSN{0001-0782, 1557-7317}


\bibitem[\protect\citeauthoryear{Richard and Kafai}{Richard and Kafai}{2016}]%
        {richard2016}
\bibfield{author}{\bibinfo{person}{G.~T. Richard} {and} \bibinfo{person}{Y.~B.
  Kafai}.} \bibinfo{year}{2016}\natexlab{}.
\newblock \showarticletitle{Blind {{Spots}} in {{Youth DIY Programming}}:
  Examining {{Diversity}} in {{Creators}}, {{Content}}, and {{Comments}} within
  the {{Scratch Online Community}}}. In \bibinfo{booktitle}{\emph{Proceedings
  of the 2016 {{CHI Conference}} on {{Human Factors}} in {{Computing Systems}}
  - {{CHI}} '16}}. \bibinfo{publisher}{{ACM Press}}, \bibinfo{address}{{Santa
  Clara, California, USA}}, \bibinfo{pages}{1473--1485}.
\newblock
\showISBNx{978-1-4503-3362-7}


\bibitem[\protect\citeauthoryear{Robertson}{Robertson}{2012}]%
        {robertson2012}
\bibfield{author}{\bibinfo{person}{J. Robertson}.}
  \bibinfo{year}{2012}\natexlab{}.
\newblock \showarticletitle{Making games in the classroom: Benefits and gender
  concerns}.
\newblock \bibinfo{journal}{\emph{Computers \& Education}}
  \bibinfo{volume}{59}, \bibinfo{number}{2} (\bibinfo{year}{2012}),
  \bibinfo{pages}{385--398}.
\newblock


\bibitem[\protect\citeauthoryear{Roque, Rusk, and Resnick}{Roque
  et~al\mbox{.}}{2016}]%
        {roque2016a}
\bibfield{author}{\bibinfo{person}{R. Roque}, \bibinfo{person}{N. Rusk}, {and}
  \bibinfo{person}{M. Resnick}.} \bibinfo{year}{2016}\natexlab{}.
\newblock \showarticletitle{Supporting {{Diverse}} and {{Creative
  Collaboration}} in the {{Scratch Online Community}}}.
\newblock In \bibinfo{booktitle}{\emph{Mass {{Collaboration}} and
  {{Education}}}}, \bibfield{editor}{\bibinfo{person}{U.~Cress},
  \bibinfo{person}{J.~Moskaliuk}, {and} \bibinfo{person}{H.~Jeong}} (Eds.).
  \bibinfo{publisher}{{Springer International Publishing}},
  \bibinfo{address}{{Cham}}, \bibinfo{pages}{241--256}.
\newblock
\showISBNx{978-3-319-13535-9 978-3-319-13536-6}


\bibitem[\protect\citeauthoryear{Rubegni, Landoni, and Jaccheri}{Rubegni
  et~al\mbox{.}}{2020}]%
        {rubegni2020}
\bibfield{author}{\bibinfo{person}{E. Rubegni}, \bibinfo{person}{M. Landoni},
  {and} \bibinfo{person}{L. Jaccheri}.} \bibinfo{year}{2020}\natexlab{}.
\newblock \showarticletitle{Design for {{Change With}} and for {{Children}}:
  How to {{Design Digital StoryTelling Tool}} to {{Raise Stereotypes
  Awareness}}}. In \bibinfo{booktitle}{\emph{Proceedings of the 2020 {{ACM
  Designing Interactive Systems Conference}}}}. \bibinfo{publisher}{{ACM}},
  \bibinfo{address}{{Eindhoven Netherlands}}, \bibinfo{pages}{505--518}.
\newblock
\showISBNx{978-1-4503-6974-9}


\bibitem[\protect\citeauthoryear{Rubio, {Romero-Zaliz}, Ma{\~n}oso, and {de
  Madrid}}{Rubio et~al\mbox{.}}{2015}]%
        {rubio2015}
\bibfield{author}{\bibinfo{person}{M.~A. Rubio}, \bibinfo{person}{R.
  {Romero-Zaliz}}, \bibinfo{person}{C. Ma{\~n}oso}, {and}
  \bibinfo{person}{A.~P. {de Madrid}}.} \bibinfo{year}{2015}\natexlab{}.
\newblock \showarticletitle{Closing the Gender Gap in an Introductory
  Programming Course}.
\newblock \bibinfo{journal}{\emph{Computers \& Education}}
  \bibinfo{volume}{82} (\bibinfo{date}{March} \bibinfo{year}{2015}),
  \bibinfo{pages}{409--420}.
\newblock
\showISSN{03601315}


\bibitem[\protect\citeauthoryear{Sullivan and Umashi~Bers}{Sullivan and
  Umashi~Bers}{2016}]%
        {sullivan2016}
\bibfield{author}{\bibinfo{person}{A. Sullivan} {and} \bibinfo{person}{M.
  Umashi~Bers}.} \bibinfo{year}{2016}\natexlab{}.
\newblock \showarticletitle{Girls, {{Boys}}, and {{Bots}}: Gender
  {{Differences}} in {{Young Children}}'s {{Performance}} on {{Robotics}} and
  {{Programming Tasks}}}.
\newblock \bibinfo{journal}{\emph{Journal of Information Technology Education:
  Innovations in Practice}}  \bibinfo{volume}{15} (\bibinfo{year}{2016}),
  \bibinfo{pages}{145--165}.
\newblock
\showISSN{2165-3151, 2165-316X}


\bibitem[\protect\citeauthoryear{Teague}{Teague}{2002}]%
        {teague2002}
\bibfield{author}{\bibinfo{person}{J. Teague}.}
  \bibinfo{year}{2002}\natexlab{}.
\newblock \showarticletitle{Women in computing: What brings them to it, what
  keeps them in it?}
\newblock \bibinfo{journal}{\emph{ACM SIGCSE Bulletin}} \bibinfo{volume}{34},
  \bibinfo{number}{2} (\bibinfo{year}{2002}), \bibinfo{pages}{147--158}.
\newblock


\bibitem[\protect\citeauthoryear{Usher and Pajares}{Usher and Pajares}{2008}]%
        {usher2008}
\bibfield{author}{\bibinfo{person}{E.~L. Usher} {and} \bibinfo{person}{F.
  Pajares}.} \bibinfo{year}{2008}\natexlab{}.
\newblock \showarticletitle{Sources of {{Self}}-{{Efficacy}} in {{School}}:
  Critical {{Review}} of the {{Literature}} and {{Future Directions}}}.
\newblock \bibinfo{journal}{\emph{Review of Educational Research}}
  \bibinfo{volume}{78}, \bibinfo{number}{4} (\bibinfo{date}{Dec.}
  \bibinfo{year}{2008}), \bibinfo{pages}{751--796}.
\newblock
\showISSN{0034-6543, 1935-1046}


\bibitem[\protect\citeauthoryear{Vrieler, Nyl{\'e}n, and Cajander}{Vrieler
  et~al\mbox{.}}{2020}]%
        {vrieler2020}
\bibfield{author}{\bibinfo{person}{T. Vrieler}, \bibinfo{person}{A. Nyl{\'e}n},
  {and} \bibinfo{person}{{\AA}. Cajander}.} \bibinfo{year}{2020}\natexlab{}.
\newblock \showarticletitle{Computer Science Club for Girls and Boys
  \textendash{} a Survey Study on Gender Differences}.
\newblock \bibinfo{journal}{\emph{Computer Science Education}}
  (\bibinfo{date}{Oct.} \bibinfo{year}{2020}), \bibinfo{pages}{1--31}.
\newblock
\showISSN{0899-3408, 1744-5175}


\bibitem[\protect\citeauthoryear{Weese, Feldhausen, and Bean}{Weese
  et~al\mbox{.}}{2016}]%
        {weese2016}
\bibfield{author}{\bibinfo{person}{J. Weese}, \bibinfo{person}{R. Feldhausen},
  {and} \bibinfo{person}{N. Bean}.} \bibinfo{year}{2016}\natexlab{}.
\newblock \showarticletitle{The {{Impact}} of {{STEM Experiences}} on {{Student
  Self}}-{{Efficacy}} in {{Computational Thinking}}}. In
  \bibinfo{booktitle}{\emph{2016 {{ASEE Annual Conference}} \& {{Exposition
  Proceedings}}}}. \bibinfo{publisher}{{ASEE Conferences}},
  \bibinfo{address}{{New Orleans, Louisiana}}, \bibinfo{pages}{26179}.
\newblock


\bibitem[\protect\citeauthoryear{Ying, Pezzullo, Ahmed, Crompton, Blanchard,
  and Boyer}{Ying et~al\mbox{.}}{2019}]%
        {ying2019}
\bibfield{author}{\bibinfo{person}{K.~M. Ying}, \bibinfo{person}{L.~G.
  Pezzullo}, \bibinfo{person}{M. Ahmed}, \bibinfo{person}{K. Crompton},
  \bibinfo{person}{J. Blanchard}, {and} \bibinfo{person}{K.~E. Boyer}.}
  \bibinfo{year}{2019}\natexlab{}.
\newblock \showarticletitle{In {{Their Own Words}}: Gender {{Differences}} in
  {{Student Perceptions}} of {{Pair Programming}}}. In
  \bibinfo{booktitle}{\emph{Proceedings of the 50th {{ACM Technical Symposium}}
  on {{Computer Science Education}}}}. \bibinfo{publisher}{{ACM}},
  \bibinfo{address}{{Minneapolis MN USA}}, \bibinfo{pages}{1053--1059}.
\newblock
\showISBNx{978-1-4503-5890-3}


\bibitem[\protect\citeauthoryear{Zeid and {El-Bahey}}{Zeid and
  {El-Bahey}}{2011}]%
        {zeid2011}
\bibfield{author}{\bibinfo{person}{A. Zeid} {and} \bibinfo{person}{R.
  {El-Bahey}}.} \bibinfo{year}{2011}\natexlab{}.
\newblock \showarticletitle{Impact of Introducing Single-Gender Classrooms in
  Higher Education on Student Achievement Levels: A Case Study in Software
  Engineering Courses in the {{GCC}} Region}. In \bibinfo{booktitle}{\emph{2011
  {{Frontiers}} in {{Education Conference}} ({{FIE}})}}.
  \bibinfo{pages}{T2H--1--T2H--6}.
\newblock
\showISSN{2377-634X}


\bibitem[\protect\citeauthoryear{Zhan, Fong, Mei, and Liang}{Zhan
  et~al\mbox{.}}{2015}]%
        {zhan2015}
\bibfield{author}{\bibinfo{person}{Z. Zhan}, \bibinfo{person}{P. Fong},
  \bibinfo{person}{H. Mei}, {and} \bibinfo{person}{T. Liang}.}
  \bibinfo{year}{2015}\natexlab{}.
\newblock \showarticletitle{Effects of Gender Grouping on Students' Group
  Performance, Individual Achievements and Attitudes in Computer-Supported
  Collaborative Learning}.
\newblock \bibinfo{journal}{\emph{Computers in Human Behavior}}
  \bibinfo{volume}{48} (\bibinfo{date}{July} \bibinfo{year}{2015}),
  \bibinfo{pages}{587--596}.
\newblock


\bibitem[\protect\citeauthoryear{Zhong, Wang, and Chen}{Zhong
  et~al\mbox{.}}{2016}]%
        {zhong2016}
\bibfield{author}{\bibinfo{person}{B. Zhong}, \bibinfo{person}{Q. Wang}, {and}
  \bibinfo{person}{J. Chen}.} \bibinfo{year}{2016}\natexlab{}.
\newblock \showarticletitle{The Impact of Social Factors on Pair Programming in
  a Primary School}.
\newblock \bibinfo{journal}{\emph{Computers in Human Behavior}}
  \bibinfo{volume}{64} (\bibinfo{date}{Nov.} \bibinfo{year}{2016}),
  \bibinfo{pages}{423--431}.
\newblock
\showISSN{0747-5632}


\end{thebibliography}

\end{document}